\documentclass[paper]{JHEP3}
\usepackage{epsfig}
\newcommand{\qu}{u}
\newcommand{\qd}{d}
\newcommand{\qubar}{\bar u}
\newcommand{\qdbar}{\bar d}
\newcommand{\bfig}{\begin{center}\begin{picture}}
\newcommand{\efig}[1]{\end{picture}\\{\small #1}\end{center}}

\newcommand{\bmip}[2]{\begin{minipage}[t]{#1pt}\bfig(#1,#2)}
\newcommand{\emip}[1]{\efig{#1}\end{minipage}}

\newcommand{\bq}{\begin{equation}}
\newcommand{\eq}{\end{equation}}
\newcommand{\bqa}{\begin{eqnarray}}
\newcommand{\eqa}{\end{eqnarray}}

\newcommand{\gev}{\mbox{GeV}}

        \newdimen\mysep                
        \newdimen\hmysep
  \newcommand{\ccaption}[2]{
    \begin{center}
    \parbox{0.85\textwidth}{
      \caption[#1]{\small{{#2}}}
      }
    \end{center}
    }

\def    \be             {\begin{equation}}
\def    \ee             {\end{equation}}
\def    \ba             {\begin{eqnarray}}
\def    \ea             {\end{eqnarray}}
\def    \nn             {\nonumber}
\def    \=              {\;=\;}
\def    \frac           #1#2{{#1 \over #2}}

\def    \bra#1          {\mbox{$\langle #1 |$}}
\def    \ket#1          {\mbox{$| #1 \rangle$}}

\def    \gev            {\mbox{$\mathrm{GeV}$}}

\def    \nubar   {\bar{\nu}}
\def    \ubar   {\bar{u}}
\def    \dbar   {\bar{d}}
\def    \qbar   {\bar{q}}
\def    \bbar   {\bar{b}}
\def    \cbar   {\bar{c}}
\def    \cpbar   {\bar{c}'}
\def    \tbar   {\bar{t}}
\def    \Qbar   {\overline{Q}}

\def    \mZ             {\mbox{$m_Z$} }
\def    \mZsq             {\mbox{$m_Z^2$} }
\def    \mW             {\mbox{$m_W$} }
\def    \mWsq             {\mbox{$m_W^2$} }
\def    \mH             {\mbox{$m_H$} }
\def    \mHsq             {\mbox{$m_H^2$} }
\def    \mt             {\mbox{$m_t$}}  
\def    \mtsq             {\mbox{$m_t^2$}}  
\def    \mb             {\mbox{$m_b$}}

\def    \pt             {\mbox{$p_T$}}
\def    \et             {\mbox{$E_T$}}

\def    \ptsq           {\mbox{$p^2_T$}}
\def    \ptbsq           {\mbox{$p^2_{T,b}$}}
\def    \ptbbsq           {\mbox{$p^2_{T,\bar{b}}$}}
\def    \ptWsq           {\mbox{$p^2_{T,W}$}}
\def    \ptZsq           {\mbox{$p^2_{T,Z}$}}
\def    \mT             {\mbox{$m_T$}}
\def    \mTsq           {\mbox{$m^2_T$}}

\newcommand     \MSB            {\ifmmode {\overline{\rm MS}} \else 
                                 $\overline{\rm MS}$  \fi}

\def    \as             {\ifmmode \alpha_s \else $\alpha_s$ \fi}
\def    \aem             {\mbox{$\alpha_{em}(\mZ)$}}

\def \oacube {\mbox{$ {\cal O}(\alpha_s^3)$}}

\def\rt1{\raisebox{-1ex}{\rlap{$\; \rho \to 1 \;\;$}}
\raisebox{.4ex}{$\;\; \;\;\simeq \;\;\;\;$}}

\def\herwig{{\small HERWIG}}
\def\isajet{{\small ISAJET}}
\def\pythia{{\small PYTHIA}}
\def\grace{{\small GRACE}}
\def\amegic{{\small AMEGIC++}}
\def\vecbos{{\small VECBOS}}
\def\madgraph{{\small MADGRAPH}}
\def\comphep{{\small CompHEP}}
\def\ALPHA{{\small ALPHA}}
\def\ppbar{\mbox{$p \bar{p}$}}
\def\met{$\rlap{\kern.2em/}E_T$}

\title{ 
      ALPGEN, a generator for 
      hard multiparton processes 
      in hadronic collisions \thanks{The work of MLM and FP is
        supported in part by the EU Fourth Framework Programme
        ``Training and Mobility of Researchers'', Network ``Quantum
        Chromodynamics and the Deep Structure of Elementary
        Particles'', contract FMRX--CT98--0194 (DG 12 -- MIHT). MM and
        RP acknowledge the financial support of the European Union
        under contract HPRN-CT-2000-00149. ADP is supported by a
        M.Curie fellowship, contract HPMF-CT-2001-01178.}}
\vfill                                                       
\author{
  Michelangelo L. MANGANO, Fulvio PICCININI\thanks{On leave 
          of absence from INFN Sezione di Pavia, Italy.},
and Antonio D. POLOSA\\
CERN, Theoretical Physics Division, CH~1211 Geneva 23, Switzerland
\\ E-mail: \email{michelangelo.mangano@cern.ch, fulvio.piccinini@cern.ch,
 antonio.polosa@cern.ch}}
\author{Mauro MORETTI\\
Dipartimento di Fisica, Universit\`{a} di 
Ferrara, and INFN, Ferrara, Italy \\
E-mail: \email{mauro.moretti@fe.infn.it}}
\author{Roberto PITTAU \\ 
Dipartimento di Fisica, 
Universit\`{a} di Torino, and INFN, Torino, Italy \\
E-mail: \email{roberto.pittau@to.infn.it}}
\abstract{ This paper presents a new event generator, ALPGEN,
  dedicated to the study of multiparton hard processes in hadronic
  collisions. The code performs, at the leading order in QCD and EW
  interactions, the calculation of the exact matrix elements for a
  large set of parton-level processes of interest in the study of the
  Tevatron and LHC data.  The current version of the code describes
  the following final states: $(W\to f\bar{f}') Q\Qbar+ N$~jets ($Q$
  being a heavy quark, and $f=\ell,q$), with $N\le 4$;
  $(Z/\gamma^{*}\to f\bar{f}) \, Q\Qbar+N$ ~jets ($f=\ell,\nu$), with
  $N\le 4$; $(W\to f\bar{f}') + \mbox{charm} + N$~jets ($f=\ell,q$,
  $N\le 5$); $(W\to f\bar{f}') + N$~jets ($f=\ell,q$) and
  $(Z/\gamma^{*}\to f\bar{f})+ N$~jets ($f=\ell,\nu$), with $N\le 6$;
  $nW+mZ+lH+N$~jets, with $n+m+l+N\le 8$, $N\le3$, including all
  2-fermion decay modes of $W$ and $Z$ bosons, with spin correlations;
  $Q\Qbar+N$~jets, with $t\to b f\bar{f}'$ decays and relative spin
  correlations included if $Q=t$, and $N\le 6$; $Q\Qbar
  Q'\Qbar'+N$~jets, with $Q$ and $Q'$ heavy quarks (possibly equal)
  and $N\le 4$; $H Q \Qbar+N$~jets, with $t\to b f\bar{f}'$ decays and
  relative spin correlations included if $Q=t$ and $N\le 4$;
  $N$~jets, with $N\le 6$.
  Parton-level events are generated, providing full information on
  their colour and flavour structure, enabling the evolution of the
  partons into fully hadronised final states.}

\preprint{CERN-TH/2002-129\\ FTN/T-2002/06\\ hep-ph/0206293}

\begin{document}
\section{Introduction}
\label{sec:intro}
The large energies available in current and forthcoming hadronic
colliders make final states with several hard and well separated jets
a rather common phenomenon.  These multijet final states can originate
directly from hard QCD radiative processes~\cite{Mangano:1991by}, or
from the decay of massive particles, such as for example $W$ and $Z$
gauge bosons, top quarks, Higgs bosons, supersymmetric particles, etc.
Whether our interest is in accurate measurements of top quarks or in
the search for more exotic states~\cite{Gianotti:2002xx}, multijet
final states always provide an important observable, and the study of
the backgrounds due to QCD is an essential part of any experimental
analysis.

Several examples of calculation of multijet cross-sections in hadronic
collisions exist in the literature. Some of them are included in
parton-level Monte Carlo (MC) event generators, where final states
consisting of hard and well isolated partons are generated. Among the
most used and best documented examples are
PAPAGENO~\cite{Hinchliffe:1993de} (a compilation of several partonic
processes), VECBOS~\cite{Berends:1991ax} (for production of $W/Z$
bosons in association with up to 4 jets), NJETS~\cite{Berends:1989ie}
(for production of up to 6 jets). The range of jet multiplicities
calculable in practice for purely QCD processes in hadronic collisions
has recently been extended in~\cite{Draggiotis:2002hm}, where new
techniques~\cite{Draggiotis:2000sh} to deal with the complexity of the
multijet flavour configurations have been implemented in the
calculation of up to 7 jets.  Finally, programs for the automatic
generation of user-specified parton-level processes exist and have
been used in the literature for the calculation of many important
reactions in hadronic collisions: \madgraph~\cite{Stelzer:1994ta},
\comphep~\cite{Pukhov:1999gg}, \grace~\cite{Ishikawa:1993qr} and
\amegic~\cite{Krauss:2001iv}.

In order to use these results in practical analyses of the
experimental data, the calculations need to be completed with the
treatment of the higher-order corrections leading to the development
of partonic cascades, and with the subsequent transformation of the
partons into observable hadrons.  MC programs such as
\herwig~\cite{Marchesini:1988cf}, \pythia~\cite{Sjostrand:1994yb} or
\isajet~\cite{Paige:1998xm} are available to carry out these last two
steps. The consistent combination of the parton-level calculations
with the partonic evolution given by the shower MC programs is the
subject of extensive work.  Several approaches to this problem have
been proposed in the case of low-order processes, where one is
interested in final states with at most one extra jet relative to a
given Born-level configuration (for example $W,Z$ plus
jet~\cite{Seymour:1995df} and $t\bar{t}$ +
jet~\cite{Corcella:1998rs}).  In these cases, the proposed algorithms
correct the probability for hard-emission estimated by the
approximation of the shower-evolution programs, using the value of the
exact real-emission higher-order matrix element. In other
studies~\cite{Dobbs:2001gb,Frixione:2002ik,Grace:2003npb,Frixione:2003ei},
covering at this time jet emission in association with DY, vector
boson pairs and heavy quark pairs, the full set of virtual and real
NLO corrections to the partonic matrix elements has been merged with
the \herwig\ MC. See in particular~\cite{Frixione:2002ik} for a
complete discussion of the problem of matching NLO parton level and
leading-logarithmic (LL) shower generators.  Some of the problems
raised by a consistent matching of NLO parton level calculations and
next-to-leading logarithm (NLL) shower evolution are discussed
in~\cite{Collins:2000qd}.

In the case of large jet multiplicities, the complexity of the matrix
element evaluation and of its singularity structure prevents so far
the application of these approaches. Recently, a new strategy has
been introduced~\cite{Krauss:1999mt,Catani:2001cc} for the merging of multijet
matrix elements with the shower development. This involves a
reweighting of the matrix element weights with Sudakov form factors,
and the veto of shower emissions in regions of phase-space already
covered by the parton-level configurations. After the shower
evolution, samples of different parton-level multiplicity are combined
together to obtain inclusive samples of arbitrary jet multiplicity,
double counting being limited to subleading effects. A complete
application of these ideas has been achieved for inclusive hadronic
final states in $e^+e^-$ collisions~\cite{Kuhn:2000dk,Catani:2001cc}, 
and a proposal has been
formulated for the extension to the hadron collider
case~\cite{Krauss:2002up}. Explicit implementations are being
developed~\cite{kraussetal}.
 
 We discussed in~\cite{Caravaglios:1999yr} a theoretical framework for
the evaluation and generation of events in such a way as to enable the
subsequent perturbative evolution using a shower MC program.
In~\cite{Mangano:2001xp} we presented a complete application to the
case where a $W$ boson is produced in association with a heavy quark
pair, plus up to 4 additional light partons. The relative MC code
allows a complete description of these final states, from the leading
order (LO) matrix element computation to the perturbative evolution
and hadronization carried out using \herwig.

We recently extended the application of the ideas contained
in~\cite{Caravaglios:1999yr}, completing a library of MC codes for
hadronic collisions including the following new processes: $WQ\Qbar+
N$~jets and $Z/\gamma^{*} \, Q\Qbar+N$ ~jets ($Q$ being a heavy
quark), with $N\le 4$; $W+\mbox{charm}+ N$~jets; 
$W+ N$~jets and $Z+ N$~jets ($N\le 6$);
$nW+mZ+lH+N$~jets, with $n+m+l+N\le 8$ and $N\le3$; $Q\Qbar+N$~jets
($N\le 6$); $Q\Qbar Q'\Qbar'+N$~jets, with $Q$ and $Q'$ heavy quarks
(possibly equal) and $N\le 4$; $H Q \Qbar+N$~jets ($N\le 4$);
$N$~jets, with $N\le 6$. Details on the decay mode options for the
various unstable particles will be  given below. For all of these
processes, an interface to both \herwig\ and \pythia\ is provided. In the
present work, we document the contents and use of this library. The
emphasis is on the code itself and not on the physics approach, which
is  discussed in more detail in~\cite{Mangano:2001xp}. Numerical results for
some benchmark processes are nevertheless presented.

Independent work on the merging of parton-level calculations with
shower MC's has been pursued by the \comphep\
group~\cite{Belyaev:2000wn}, by the \grace\ group~\cite{Sato:2001ae}
and, more recently,
in~\cite{Kersevan:2002dd,Tsuno:2002ae,Maltoni:2002qb}.  To the best of
our knowledge, a large fraction of the matrix element calculations
documented in this paper have however never been performed before.

Section~\ref{sec:general} reviews the general structure of the codes,
covering both the aspects of the  parton-level calculations, and
of the shower evolution.  Section~\ref{sec:hard} discusses the
features of the implementation of each individual hard process in our
library.  A technical Appendix will provide more explicit details on
the programs and their use.

The library containing all codes described in this work can be
downloaded from the following URL: {\tt
  http://mlm.home.cern.ch/mlm/alpgen} \,.

\section{The general structure of the program}
\label{sec:general}
The program consists of several building blocks (see the Appendix for
details). A section of the code library defines the overall
infrastructure of the generator, implementing the logical sequence of
operations in a standard set of subroutine calls; this part is
independent of the hard process selected and, among other things, it
includes the algorithms needed for the evaluation of the matrix
elements, for the evaluation of the parton densities and for the
bookkeeping and histogramming of the results.  Each hard process has
viceversa a separate set of code elements, which are specific to it.
These include the process initialization, the phase-space generation,
the extraction of flavour and colour structure of the event, and the
default analysis routines.  Each hard process corresponds to a
specific executable, obtained by linking the relative
process-dependent code elements with the process-independent ones.  In
addition to the above, a section of the code library deals with the
shower evolution. As explained below, the shower evolution is
performed as an independent step, following the generation of a sample
of unweighted parton-level events. The code elements relative to this
phase of the computation include the \herwig\ and \pythia\ codes and
the algorithm needed to transform the partonic input into a format
which can be interpreted and processed by \herwig\ and \pythia. We adopt the
formatting standard proposed in the so called Les Houches
accord~\cite{Boos:2001cv}.

As alluded to above, the program has two main modes of operation. In
the first mode the code performs the parton-level calculation of the
matrix elements relative to the selected hard processes, generating
weighted events.  Each weighted event is analysed on-line in a routine
where the kinematics of the event can be studied, and histograms
filled. The user has direct access to this analysis routine, and can
adapt it to his needs.  At the end of the run, differential
distributions are obtained. A histogramming package is included in the
code library, which generates a graphic output in the form of {\tt
topdrawer}~\cite{topdrawer} plots.  This mode of running can also be
used to easily get total cross sections in presence of some
overall generation cuts (e.g. rates for production of jets above a
given threshold), without looking at any particular differential
distribution.

In the second mode of operation the code generates unweighted parton
level events and stores them to a file, for subsequent evolution via
the parton-shower part of the program. The generation of parton-level
events, and their shower evolution, are performed in two different
phases by different code elements. Since the generation of unweighted
parton-level events is typically by far the most CPU intensive
component of the calculation, the storage of unweighted events allows
to build up event data sets which can then be used efficiently for
studies of hadronization systematics or realistic detector
simulations. In this mode of operation the matrix-element calculation
generates all the flavour and colour information necessary for the
complete shower evolution. The kinematical,
flavour and colour data for a given event are stored in a file, and
are read in by the shower MC, which will process the event.
In the rest of this Section we present in more detail these two running
modes of the code.

\subsection{Parton-level generation and cross section evaluation}
\label{sec:xsec}
In a nutshell, the calculation of the cross section for a given hard
process is performed in the following steps:
\begin{itemize}
\item The parameters required to define the hard process are passed to
  the code. These include the selection of jet multiplicity, the
  mass of possible heavy quarks, rapidity and transverse momentum
  cuts, etc.
\item A first set of phase-space integration cycles is performed, with
 the goal of exploring how the cross-section is distributed in
 phase-space and among the possible contributing subprocesses. Event
 by event, the following steps take place:
\begin{itemize}
  \item one subprocess (see later) is randomly selected; 
  \item a point in phase-space is randomly selected, consistent with
    the required kinematical acceptance cuts;
  \item the initial-state parton luminosity is evaluated for the
    chosen subprocess, and 
   one among the possible flavour configurations is selected (see later);
  \item
    spin and colour for each parton are randomly assigned;
  \item the matrix element is evaluated, and the weight of the event
    is obtained after inclusion of the phase-space and parton-luminosity
    factors. A bookkeeping of the weights is kept for each individual
   subprocess and phase-space subvolume.
\end{itemize}
\item At the end of the first integration iteration, a map of the
  cross-section distribution among the different subprocesses and in
  phase-space is available. It will be used for subsequent
  integration cycles, where the phase-space and subprocess random sampling
  will be weighted by the respective probability distributions.
\item After the completion of a series of warm-up integration cycles (whose
  number is specified at the beginning of the run by the user), the
  optimised integration grids are stored in a file. A final
  large-statistics run is then performed. After the generation of each
  event, its kinematics is analysed and histograms are filled.
\end{itemize}
We shall now discuss in more detail the individual steps outlined
above and the ingredients of the calculation.

\subsubsection{Selection of the subprocess}
The calculation of the cross section for  multiparton final
states involves typically the sum over a large set of subprocesses
and flavour configurations. For example, in the case of $WQ\Qbar+2$
jets we have, among others, the following subprocesses:
\be q\qbar' \to W Q\Qbar g g, \;
q g \to W Q\Qbar g  q' , \;
g q \to  W Q\Qbar g  q' , \;
gg \to W Q\Qbar q \qbar' , \;
q\qbar' \to W Q \Qbar q'' \qbar''  , \; 
{\mbox{ etc.}}
\ee
Each of these subprocesses receives contributions from several
possible flavour configurations (e.g. $u\bar{d} \to W Q \Qbar gg$ ,
$u\bar{s} \to W Q \Qbar gg$, etc.). Our subdivision in subprocesses is
designed to allow to sum the contribution of different flavour
configurations by simply adding trivial factors such as parton
densities or CKM factors, which factorize out of a single, flavour
independent,  matrix element. For example the overall
contribution from the first process in the above list is given by 
\be \label{eq:ckm}
\left [ u_1\bar{d}_2\cos^2\theta_c + u_1\bar{s}_2\sin^2\theta_c +
  c_1\bar{s}_2\cos^2\theta_c + c_1\bar{d}_2\sin^2\theta_c \right ]
\times \vert M(q\qbar' \to W Q\Qbar g g)\vert^2 \; , 
\ee 
where $q_i=f(x_i),\; i=1,2$,  are the parton densities for the
quark flavour $q$. 
Contributions from
charge-conjugate or isospin-rotated states can also be summed up,
after trivial momentum exchanges. 
For example, the same matrix element
calculation is used to describe  the four events:
\ba
u(p_1) \dbar(p_2) &\to& b(p_3) \bbar(p_4) g(p_5) g(p_6) e^+(p_5)
\nu(p_6)
\nn \\
\ubar(p_1) d(p_2) &\to& \bbar(p_3) b(p_4) g(p_5) g(p_6) e^-(p_5)
\nubar(p_6)
\nn \\
\dbar(p_1) u(p_2) &\to& \bbar(p_3) b(p_4) g(p_5) g(p_6) \nu(p_5)
e^+(p_6)
\nn \\
d(p_1) \ubar(p_2) &\to& b(p_3) \bbar(p_4) g(p_5) g(p_6) \nubar(p_5)
e^-(p_6) \; . \nn
\ea
Event by event, the flavour configuration for the
assigned subprocess is then selected with a probability
proportional to the relative size of the individual contributions to
the luminosity, weighted by the Cabibbo angles. 

Typically, only few among all possible subprocesses give a substantial
contribution to the cross section. For each event, instead of summing
the weight of all subprocesses, we calculate only one. This is
selected with uniform probability during the first integration cycle,
and a record is kept of each individual
contribution to the cross-section. In subsequent integration
iterations, the accumulated rates of the single channels are used to
weight their selection probabilities. This will significantly improve
the CPU performance of the code, and the unweighting efficiency.

Due to the rapid growth in the number of subprocesses when quarks are
added~\cite{Draggiotis:2000sh}, we limited ourselves to processes with
at most 2 pairs of light quarks (plus pairs of heavy quarks, when
required). In all of the cases considered this is not, however, a
significant limitation to the accuracy of the results.  The full list
of subprocesses available for each hard process is given in
Section~\ref{sec:hard}.

\subsubsection{Phase-space sampling}
The phase-space generation is optimised for each individual hard
process, using generation variables which are most suitable to the
application of typical hadron collider selection cuts.  The
phase-space is mapped with a multidimensional grid, and during the
integration a record is kept of the total weight accumulated within
each bin of the grid.  To further contribute to the efficiency of the
phase-space sampling, independent grids are employed to sample
different subprocesses. In particular, we shall associate one
phase-space grid to each of the following initial states:
\begin{enumerate}
\item $q\qbar$, $q\qbar'$ and charge conjugates
\item $qg$ and $\qbar g$
\item $gq$ and $g \qbar$
\item $gg$
\item $qq$, $qq'$ and  charge conjugates.
\end{enumerate}
For example, the processes $q \qbar'  \to WQ\Qbar  gg$ and 
 $q \qbar'  \to WQ\Qbar  q \qbar$ share the same phase-space
 grid.
 
\subsubsection{Matrix element calculation}
The calculation of the LO matrix elements for the selected hard
process is performed using the \ALPHA\cite{Caravaglios:1995cd}
algorithm, extended to QCD interactions as described
in~\cite{Caravaglios:1999yr,Mangano:2001xp}.  As explained in detail
in~\cite{Mangano:2001xp}, use of the \ALPHA\ algorithm is in our view
essential in order to cope with the complexity of the problem.  All
mass effects are included in the case of massive quarks.  The
calculations are done after having selected, on an event-by-event
basis, polarization, flavour and colour configurations, in order to be
able to provide the shower MC's with the information necessary for the
shower evolution. The sum over polarizations and colours is performed
by summing over multiple events, in a MC fashion.  The choice of
colour basis and the strategy for the determination of the colour
flows necessary for the coherent shower evolution are described
in~\cite{Mangano:2001xp}. 
To the best of our knowledge, a large fraction of the matrix element
calculations documented in this paper have never been performed before
using other calculational tools.  An independent implementation of the
\ALPHA\ algorithm has been exploited for the evaluation of multijet
processes in hadronic collisions
in~\cite{Draggiotis:1998gr,Draggiotis:2002hm}.
 
\subsubsection{PDF sets and  $\mathbf{\alpha_s}$}
 The code library includes a choice among some of the most recent PDF
 parameterizations. They can be selected at the beginning of the run
 through the variable {\tt ndns}, which is  mapped as
 follows\footnote{In addition to these default sets, we
 include in the package the full group of 40 CTEQ6M sets which allow
 the determination of PDF systematics errors~\cite{Pumplin:2002vw}. 
 To access these, the file {\tt alplib/alppdf.f} should be
 replaced in the linking stage with the file  {\tt
 alplib/alppdf}$\_${\tt cteq.f}.
 A more complete set of parton densities, including
   old sets back to EHLQ and DO, is also available upon request.}:
\def\cteqa {1 & CTEQ4M~\cite{Lai:1996mg} & $[0.116]_2$}
\def\cteqb {2& CTEQ4L~\cite{Lai:1996mg}    &  $[0.116]_2$ }
\def\cteqc {3& CTEQ4HJ~\cite{Lai:1996mg}   & $[0.116]_2$ } 
\def\cteqd {4& CTEQ5M~\cite{Lai:2000wy}    &  $[0.118]_2$ }
\def\cteqe {5& CTEQ5L~\cite{Lai:2000wy}    & $[0.127]_1$ }
\def\cteqf {6& CTEQ5HJ~\cite{Lai:2000wy}   & $[0.118]_2$ } 
\def\cteqg {7& CTEQ6M~\cite{Pumplin:2002vw}& $[0.118]_2$  }
\def\cteqh {8& CTEQ6L~\cite{Pumplin:2002vw} & $[0.118]_2$  }
\def\mrsa{101 & MRST99-1~\cite{Martin:1999ww} & $[0.1175]_2$ }
\def\mrsb{102 & MRST01-1~\cite{Martin:2001es}  & $[0.119]_2$ } 
\def\mrsc{103 & MRST01-2~\cite{Martin:2001es} & $[0.117]_2$ } 
\def\mrsd{104 & MRST01-3~\cite{Martin:2001es} & $[0.121]_2$ } 
\def\mrse{105 & MRST01J~\cite{Martin:2001es} & $[0.121]_2$ } 
\def\mrsf{106 & MRSTLO~\cite{Martin:2002dr} & $[0.130]_1$ }
\def\dummypdf{& & }
\begin{center}
{\small
\begin{tabular}{lll|lll}
{\tt ndns} & PDF & $[\as(\mZ)]_{n_{\rm loop}}$ &
{\tt ndns} & PDF & $[\as(\mZ)]_{n_{\rm loop}}$ \\ \hline
\cteqa & \mrsa \\
\cteqb & \mrsb \\
\cteqc & \mrsc \\
\cteqd & \mrsd \\
\cteqe & \mrse \\
\cteqf & \mrsf \\
\cteqg & \dummypdf \\
\cteqh & \dummypdf \\
\end{tabular}
}
\end{center}
In the case of NLO sets we use the 2-loop expression for $\as$: 
\be
\as(Q)=\frac{1}{b_5 \, \log(Q^2/\Lambda_5^2)} - \frac{b'_5}{(b_5 \,
  \log^2(Q^2/\Lambda_5^2))\log\log(Q^2/\Lambda_5^2)} \; .
\ee 
valid for $Q>m_b\equiv4.5$~GeV, 
where $b_5$ and $b'_5$ are the 1- and 2-loop coefficients of the QCD
$\beta$ function, respectively, for 5 flavours. Threshold matching is
applied in the case of $Q<m_b$. In the case of LO sets (such as
CTEQ*L or MRSTLO) we follow the prescriptions used by the authors in
performing the PDF fits. These vary from set to set. For example,  set
CTEQ5L was fitted using a LO expression for $\as$, while CTEQ6L used
the NLO evolution. The table above lists the values of $\as(\mZ)$
corresponding to the various sets, and indicates whether these values
(and the relative evolution to different renormalization scales)
correspond to the 1 or 2 loop formulation.

\subsubsection{Electroweak couplings}
In the current version of the \ALPHA\ code the input couplings are
derived from the standard $SU(3)\times SU(2)_L \times U(1)_Y$ tree
level Lagrangian.  The choice of input EW parameters deserves a short
discussion. In \ALPHA, the couplings of the electroweak (EW) and Higgs
($H$) bosons (including the respective selfcouplings) are parametrised
in terms of the $SU(2)$ coupling strength $g$, of the weak mixing
angle $\sin\theta_W$, of the electromagnetic fine structure constant
$\alpha_{em}$, and of the masses of $W$, $Z$ and $H$. We therefore
have a total of 6 parameters needed to specify the value of general EW
matrix elements. If we want to preserve gauge invariance we are
however allowed to use only four independent parameters (plus fermion
masses in the Yukawa sector). Treating $\mH$ as a free parameter,
gauge invariance at the tree level demands that the remaining 5
parameters satisfy the following tree-level relationships: \ba
\cos \theta_W & = & \frac {\mW} {\mZ} \\
e &=& g \; \sin\theta_W \; . \ea The Higgs self-couplings and Yukawa
couplings to fermions are furthermore given by: \ba
\lambda_{hhh} & = & \frac{g \, \mHsq}{4\mW} \\
\lambda_{hhhh} & = & \frac{g^2 \, \mHsq}{32 \, M^2_W} \\
y_f & = & \frac{gm_f}{\sqrt{2}\mW} \; . \ea As a consequence, we cannot
assign to the input parameters the values which are known from the
current accurate experimental measurements, since these values are
only consistent with the radiatively corrected versions of the above
relations.  This is not a merely formal issue: any tiny violation of
the tree-level gauge relationships among the model parameters leads to
violations of the equivalence theorem and leads to unphysical
corrections to the tree-level cross-sections scaling like
$(E_V/M_V)^{2n}$, $n$ being the number of on/off-shell heavy gauge
bosons appearing in the relevant diagrams and $E_V$, $M_V$ their
energy and mass respectively.  In view of the large center of mass
energy available in current and future hadron collisions, these
spurious corrections could be numerically large, leading to the wrong
high-energy behaviour of the cross sections.

The current version of the code provides four choices for the setting
of EW parameters. These choices are controlled by the variable {\tt
  iewopt}, set at running time (default values for this variable are
provided in the code, and are listed in the following sections
describing the individual hard processes). The different options are
listed here below;  the numerical values of the calculated parameters are
obtained using the following set of inputs: $\mW=80.41$, $\mZ=91.188$,
$\sin^2\theta_W=0.231$, $\aem=1/128.89$, $G_F=1.16639\times 10^{-5}$ :
\begin{itemize}
\item[~] {\tt iewopt=0}.  Inputs: $\aem$, $G_{F}$,
  $\sin^2 \theta_W$. Extracted:
  \ba &&  g=\sqrt{4\pi\aem}/\sin\theta_W = 0.6497, \quad
         \mW=g/\sqrt{4\sqrt{2} G_F} = 79.98, \label{eq:iew0} \\
      &&   \mZ = \mW/\cos\theta_W = 91.20 \nn
  \ea
\item[~] {\tt iewopt=1}.  Inputs: $\mW$, $G_{F}$,
  $\sin^2 \theta_W$. Extracted:
  \ba && \mZ=\mW/\cos\theta_W = 91.695, \quad 
  g= ( 4\sqrt{2} \, G_F)^{1/2} \mW =0.6532,  \label{eq:iew1} \\ 
     && \aem = (g \; \sin\theta_W)^2/4\pi = 1/127.51 \nn 
  \ea
\item[~] {\tt iewopt=2}.  Inputs: $\mZ$, $\aem$,
  $\sin^2 \theta_W$. Extracted:
  \be \mW=\mZ\cos\theta_W = 79.97 , \quad 
  g=\sqrt{4\pi\aem}/\sin\theta_W = 0.6497  \label{eq:iew2} 
  \ee
\item[~] {\tt iewopt=3}.  Inputs: $\mZ$, $\mW$,
  $G_F$. Extracted:
  \ba && \sin^2\theta_W = 1- (\mW/\mZ)^2 = 0.2224 , \quad 
  g= ( 4\sqrt{2} \, G_F)^{1/2} \mW =0.653,  \label{eq:iew3} \\
      &&  \aem = (g \; \sin\theta_W)^2/4\pi = 1/132.42 \nn \; .
  \ea
\end{itemize}
As a default, in all processes we employ {\tt iewopt=3}. We verified
that alternative options generate changes in the rates by at most few percent.
Gauge and Higgs boson widths are calculated at tree level after the
couplings have been selected. With the exception of the class of
processes involving several gauge bosons, which will be discussed in detail
Section~\ref{sec:nw} and where we set boson widths to 0,  \ALPHA\  
uses fixed widths in the propagators.

\subsection{Unweighting and Shower evolution}
\label{sec:shower}
The starting point for the processing of events through the shower
evolution is the generation of a sample of unweighted events.
This generation takes place through a two-step procedure (more details
are given in the Appendix):
\begin{enumerate}
\item to start with, a run of the parton-level code is performed as
  described in Section~\ref{sec:xsec}. Selecting the running mode option
  {\tt imode=1}, weighted events are stored in a file. To limit the
  size of the file, instead of saving all the event information, we
  simply store the seed of the first random number used in the
  generation of the event, in addition to the event weight.
\item at the end of the generation of the weighted event sample, the
  unweighting is performed by running the code once more, using a
  running mode option {\tt imode=2}. In this running mode, the code
  will sequentially read the events stored in the file, and will
  perform the unweighting using the knowledge of the maximum weight of
  the sample, and of the weight of each individual event. When an
  event is selected by the unweighting, the seed of the random number
  is uploaded and all the information about the event (kinematics,
  flavours, spins and colours) is automatically reconstructed. The
  colour flow for the event is then selected according to the
  algorithm described in~\cite{Mangano:2001xp}: the subamplitudes
  corresponding to all colour flows compatible with the colour state
  of the event are first evaluated using \ALPHA; one of them
  is then randomly extracted with a probability proportional to the
  squared modulus of the relative subamplitude. The momenta of the
  particles, together with their flavour and with the colour flow
  information, are then written to a file, which at the end of the run
  will contain the complete sample of unweighted events.
\end{enumerate}
At this point we are ready to process the events through the shower
evolution. The stored events can be read by the chosen shower MC, the
kinematics, colour and flavour information for each event being
translated into the event format established by the Les Houches
convention~\cite{Boos:2001cv}.

\section{The available hard processes}
\label{sec:hard}

\subsection{$\mathbf{WQ\Qbar+}$~jets}
\label{sec:wqq}
The physics content of the  $WQ\Qbar+$~jets code has been described in detail
in~\cite{Mangano:2001xp}, where some phenomenological applications are
also presented. We use the notation $W$ as a short hand; what is
actually calculated is the matrix element for a fermion-antifermion final
state. All spin correlations and finite width effects are therefore
accounted for.  
The quoted cross sections refer to a single lepton
family; in the flavour assignment, the code selects by default an
electron. Different flavours can be selected during the unweighting
phase, covering all possible leptonic decays, as well as inclusive
quark decays (for more details see the Appendix B.4).
In the case $Q=t$, the top quark is left undecayed.
The EW parameters are fixed by default using the option {\tt iewopt=3}
(see eq.~(\ref{eq:iew1})). Only the leading-order EW diagrams are
included in the calculation.

 The subprocesses considered include all configurations
with up to 2 light-quark pairs; they are listed in
Table~\ref{tab:wqq}, following the notation employed in the
code.
\begin{table}
\begin{center}
\begin{tabular}{ll|ll|ll}
{\tt jproc} & subprocess & {\tt jproc} & subprocess & {\tt jproc} &
subprocess \\ 
1 &  $q\qbar' \to W Q\Qbar$ 
&2 &  $q g \to q' W Q\Qbar$ 
&3 &  $g q \to q' W Q\Qbar$ 
\\
4 &  $gg \to q \qbar' W Q\Qbar$ 
&5 &  $q\qbar' \to W Q \Qbar q'' \qbar'' $ 
&6 &  $qq'' \to W Q \Qbar q' q'' $ 
\\
7 &  $q'' q \to W Q \Qbar q' q'' $ 
&8 &  $q\qbar \to W Q \Qbar q' \qbar'' $ 
&9 &  $q\qbar' \to W Q \Qbar q \qbar $ 
\\
10 &  $\qbar' q\to W Q \Qbar q \qbar $ 
&11 &  $q\qbar \to W Q \Qbar q \qbar' $ 
&12 &  $q\qbar \to W Q \Qbar q' \qbar $ 
\\
13 &  $q q \to W Q \Qbar q q' $ 
&14 &  $q q' \to W Q \Qbar q q $ 
&15 &  $q q' \to W Q \Qbar q' q' $ 
\\
16 &  $q g \to W Q \Qbar q' q''\qbar'' $ 
&17 &  $g q \to W Q \Qbar q' q''\qbar'' $ 
&18 &  $q g \to W Q \Qbar q q \qbar' $ 
\\
19 &  $q g \to W Q \Qbar q' q \qbar $ 
&20 &  $g q \to W Q \Qbar q q \qbar' $ 
&21 &  $g q \to W Q \Qbar q' q \qbar $ 
\\
22 &  $q g \to W Q \Qbar q' q' \qbar' $ 
&23 &  $g q \to W Q \Qbar q' q' \qbar' $ 
&24 &  $g g \to W Q \Qbar q \qbar' q'' \qbar'' $ 
\\
25 &  $g g \to W Q \Qbar q \qbar q \qbar' $ 
& & & &
\end{tabular}
\ccaption{}{\label{tab:wqq} Subprocesses included in the $WQ\Qbar+$jets
  code. Additional final-state gluons are not explicitly 
  shown here but are included in the code if the requested light-jet
  multiplicity ($N\le 4$) exceeds the number of indicated final-state partons.
  For example, the subprocess {\tt jproc=1} in the case of 2 light jets
  will correspond to the final state  $q\qbar' \to W Q\Qbar g g$.
  The details can be found in the subroutine {\tt selflav} of
  the file {\tt wqqlib/wqq.f}.}
\end{center}
\end{table}
For all processes, the charge-conjugate ones are always understood.
The above list fully covers all the possible processes with up to 3
light jets in addition to the heavy quarks. In the case of 4 extra
jets, we do not calculate processes with 3 light-quark pairs. Within
the uncertainties of the LO approximation, these can be safely
neglected~\cite{Berends:1991ax}.

As a default, the code generates kinematical configurations defined by
cuts applied to the following variables (the cuts related to the heavy
quarks are only applied in the case of $b$, while top quarks are
always generated without cuts):
\begin{itemize}
\item $\pt^{\rm jet}$, $\eta^{\rm jet}$, $\Delta R_{jj}$
\item $\pt^{ b}$, $\eta^{ b}$, $\Delta R_{b\bbar}$ 
\item $\pt^{\rm \ell}$, $\eta^{\rm \ell}$, $\pt^{\rm \nu}$, $\Delta
  R_{\ell j}$ ,
\end{itemize}
where $\Delta R_{ab}=\sqrt{ [(\eta_a - \eta_b)^2+(\phi_a-\phi_b)^2 ]}$.
The  cut values can be provided by the user at run
time. Additional cuts can be supplied by the user in the 
routine {\tt usrcut} contained in the user file {\tt wqqwork/wqqusr.f}.

In the code initialization phase, 
the user can select among 3 continuous choices for the parametrization
of the factorization and renormalization scale $Q$: a real input
parameter ({\tt qfac}) allows to vary the overall scale of $Q$,
$Q={\tt qfac}\times Q_0$, while the preferred functional form for
$Q_0$ is selected through the integer input parameter {\tt
  iqopt}:
{\renewcommand{\arraystretch}{1.2}
\begin{center}
\begin{tabular}{l||l|l|l}
{\tt iqopt} & 0 & 1 & 2 \\  \hline
$Q_0^2$ & $m_W^2+\pt_W^2$ & $m_W^2$ & $m_W^2+ \sum \mTsq$ 
\end{tabular}
\end{center}
}
where $\mT$ is the transverse mass defined as $\mTsq=m^2+\ptsq$,
and the sum $\sum \mTsq $ extends to all final
state partons (including the heavy quarks, excluding the $W$ decay products).

Some numerical benchmark results are given in
Table~\ref{tab:wbbxs} and ~\ref{tab:wttxs}. 
In the case of the minimal jet multiplicity, the results agree with
previous calculations (see e.g.~\cite{Kunszt:1984ri,Mangano:1993kp}).
The following scale and cuts are used:
\ba \label{eq:wqq1}
&& Q^2 = \mWsq + \ptWsq,
\\
        && \pt^{\rm jet}>20~\gev, \quad \vert \eta_j\vert < 2.5, \quad \Delta
        R_{jj} >0.7
\\
        && \pt^{b}>20~\gev, \quad \vert \eta_b \vert < 2.5, \quad \Delta
        R_{b\bbar} >0.7, \Delta R_{bj} >0.7 \; .
\label{eq:wqq2}
\ea
Here and in the following Sections we shall use
the PDF set CTEQ5L. The quoted errors reflect the statistical accuracy
of the integrations. We never tried to go beyond the percent level, to
concentrate the CPU resources on the most computationally demanding channels.
Results for the FNAL Tevatron refer to \ppbar\
collisions at $\sqrt{S}=2$~TeV, those for the LHC refer to $pp$
collisions at $\sqrt{S}=14$~TeV.

{\renewcommand{\arraystretch}{1.2}
\begin{table}
\begin{center}
\begin{tabular}{||l|l|l|l|l|l||}\hline
 & $N = 0$  & $N = 1$ & $N = 2$ & $N = 3$ & $N = 4$
                 \\  \hline
LHC (pb)   & 2.222(4) & 3.013(9) & 1.83(1) & 0.831(8) & 0.307(5)
\\ \hline
FNAL (fb)  &  332.2(7) & 86.2(4) & 18.3(2) & 3.17(3) & 0.44(3)
\\ \hline
\end{tabular}            
\ccaption{}{\label{tab:wbbxs} $\sigma(b \bbar \ell \nu + N~{\rm jets})$
at the Tevatron and 
at the LHC. Parameters and cuts are given
in eqs.~(\ref{eq:wqq1}-\ref{eq:wqq2}). }
\end{center}
\end{table}}

{\renewcommand{\arraystretch}{1.2}
\begin{table}
\begin{center}
\begin{tabular}{||l|l||}\hline
LHC (fb)   & 61.1(4)
\\ \hline
FNAL (fb)  &  1.55(1)
\\ \hline
\end{tabular}            
\ccaption{}{\label{tab:wttxs} $\sigma(t \tbar \ell \nu)$
at the Tevatron and 
at the LHC.}
\end{center}
\end{table}}

\subsection{$\mathbf{Z/\gamma^{*} \, Q\Qbar+}$~jets}
\label{sec:zqq}
We use the notation $Z/\gamma^{*}$ as a short hand; what is actually
calculated is the matrix element for a lepton-pair final state. All
spin correlations and finite width effects are therefore accounted
for. When the final state $\ell^+ \ell^-$ is selected, the
interference between intermediate $Z$ and $\gamma^{*}$ is also included.
The quoted cross sections refer to a single lepton family; in the
flavour assignement, the code selects by default the $e^+ e^-$ pair.
In the case of the final state $\nu \bar\nu$ the quoted cross sections
include the decays to all 3 neutrino flavours, although we always
label the neutrinos as $\nu_e$.  In the case $Q=t$, the top quark is
left undecayed.
The EW parameters are fixed by default using the option {\tt iewopt=3}
(see eq.~(\ref{eq:iew2})).

 The subprocesses considered include all configurations
with up to 2 light-quark pairs; they are listed in
Table~\ref{tab:zqq}, following the notation employed in the
code.
\begin{table}
\begin{center}
\begin{tabular}{ll|ll|ll}
{\tt jproc} & subprocess & {\tt jproc} & subprocess & {\tt jproc} &
subprocess \\ 
1 &  $u\ubar \to Z Q\Qbar$ &  
2 &  $d \dbar  \to Z Q\Qbar$ &  
3 &  $g g \to Z Q\Qbar$ \\
4 &  $g u  \to u Z Q\Qbar$ &  
5 &  $g d  \to d Z Q\Qbar$ &  
6 &  $ ug  \to u Z Q\Qbar$ \\
7 &  $ dg  \to d Z Q\Qbar$ &  
8 &  $g g \to u \ubar Z Q\Qbar$ &  
9 &  $g g \to d \dbar Z Q\Qbar$ \\
10 &  $u\ubar \to u \ubar Z Q\Qbar$ &  
11 &  $d\dbar \to d \dbar Z Q\Qbar$ &  
12 &  $uu     \to u u     Z Q\Qbar$ \\  
13 &  $dd     \to d d     Z Q\Qbar$ &  
14 &  $u\ubar \to u'\ubar' Z Q\Qbar$ &  
15 &  $d\dbar \to d' \dbar' Z Q\Qbar$ \\  
16 &  $uu'     \to u u'     Z Q\Qbar$ &  
17 &  $u\ubar' \to u\ubar' Z Q\Qbar$ &  
18 &  $dd'     \to dd'     Z Q\Qbar$ \\  
19 &  $d\dbar' \to d\dbar' Z Q\Qbar$ &  
20 &  $u\ubar \to d\dbar Z Q\Qbar$ &  
21 &  $d\dbar \to u\ubar Z Q\Qbar$ \\  
22 &  $ud     \to u d     Z Q\Qbar$ &  
23 &  $du     \to  du     Z Q\Qbar$ &  
24 &  $u\dbar \to u\dbar Z Q\Qbar$ \\  
25 &  $d\ubar \to d\ubar Z Q\Qbar$ &  
26 &  $u\ubar \to b\bbar Z Q\Qbar$ &  
27 &  $d\dbar \to b\bbar Z Q\Qbar$ \\  
28 &  $gg     \to b\bbar Z Q\Qbar$     & 
29 &  $gu     \to u u \ubar Z  Q\Qbar $ &
30 &  $ug     \to u u \ubar Z   Q\Qbar$      \\
31 &  $gu     \to  u u' \ubar' Z    Q\Qbar $ &
32 &  $ug     \to u u' \ubar' Z  Q\Qbar $   &
33 &  $gu     \to u d \dbar Z   Q\Qbar$    \\
34 &  $ug     \to u d \dbar Z  Q\Qbar $    &
35 &  $gu     \to u b \bbar Z  Q\Qbar $    &
36 &  $ug     \to u b \bbar Z  Q\Qbar $    \\
37 &  $gd     \to d d \dbar Z  Q\Qbar $    &
38 &  $dg     \to d d \dbar Z  Q\Qbar $    &
39 &  $gd     \to d d' \dbar' Z  Q\Qbar$   \\
40 &  $dg     \to d d' \dbar' Z  Q\Qbar $  &
41 &  $gd     \to d u \ubar Z  Q\Qbar $    &
42 &  $dg     \to d u \ubar Z  Q\Qbar $    \\
43 &  $gd     \to d b \bbar Z Q\Qbar  $   &
44 &  $dg     \to d b \bbar Z  Q\Qbar $   & & 
\end{tabular}
\ccaption{}{\label{tab:zqq} Subprocesses included in the $Z/\gamma^*
  Q\Qbar+$jets code. It is always understood that quarks $u$ and $d$
  represent generic light quarks of type up or down. The $Z$ in the table
  stands for a neutral $\ell^+ \ell^-$ ($\nu \bar\nu$) lepton pair.
  Additional final-state gluons are not explicitly shown here but are
  included in the code if the requested light-jet multiplicity ($N\le
  4$) exceeds the number of indicated final-state partons.  For
  example, the subprocess {\tt jproc=1} in the case of 2 light jets
  will correspond to the final state $u\ubar \to Z Q\Qbar g g$.  The
  details can be found in the subroutine {\tt selflav} of the file
  {\tt zqqlib/zqq.f}.}
\end{center}
\end{table}
For each process, the charge-conjugate ones are always understood.
The above list fully covers all the possible processes with up to 3
light jets in addition to the heavy quarks. In the case of 4 or more
extra jets, we do not calculate processes with 3 light-quark pairs. As
in the case of associated production with a $W$, we expect that,
within the uncertainties of the LO approximation, these can be safely
neglected.

As a default, the code generates kinematical configurations defined by
cuts applied to the following variables (the cuts related to the heavy
quarks are only applied in the case of $b$, while top quarks are
always generated without cuts):
\begin{itemize}
\item $\pt^{\rm jet}$, $\eta^{\rm jet}$, $\Delta R_{jj}$
\item $\pt^{\rm Q}$, $\eta^{\rm Q}$, $\Delta R_{Q\Qbar}$ 
\item $\pt^{\ell}$, $\eta^{\ell}$, $\Delta R_{\ell j}$, $m(\ell^+\ell^-)$, 
for $\ell^+ \ell^-$ final states
\item missing $E_T$ for $\nu \bar\nu$ final states.  
\end{itemize}
The cut on the dilepton invariant mass allows to optimise the sampling
of the DY mass spectrum if the user is interested in events off the
$Z$ peak. Additional cuts can be supplied by the user in an
appropriate routine.  The choice of factorization and renormalization
scale is similar to what given for the $WQ\Qbar$ processes, with the
$W$ mass replaced by the mass of the DY pair (if the DY mass range
excludes the value of $\mZ$), or by $\mZ$ (when the 
DY mass range includes $\mZ$). 

Some benchmark results are given in Table~\ref{tab:zqqxs}, obtained
for the following set of cuts and scale choice:
\ba \label{eq:zqq1}
&& Q^2 = \mZsq + \ptZsq,
\quad
80~\gev \leq m_{ll} \leq 100~\gev
\\
        && \pt^{\rm jet}>20~\gev, \quad \vert \eta_j\vert < 2.5, \quad \Delta
        R_{jj} >0.7
\\
        && \pt^{b}>20~\gev, \quad \vert \eta_b \vert < 2.5, \quad \Delta
        R_{b\bbar} >0.7, R_{bj} >0.7 \; .
\label{eq:zqq2}
\ea
{\renewcommand{\arraystretch}{1.2}
\begin{table}
\begin{center}
\begin{tabular}{||l|l|l|l|l|l||}\hline
 & $N = 0$  & $N = 1$ & $N = 2$ & $N = 3$ & $N = 4$\\ 
\hline
LHC, (fb) & 1840(5)  & 1085(4) & 444(3) &  154(2) & 44(1) \\ 
\hline
FNAL, (fb) & 49.3(1) & 13.18(5) & 2.57(2) & 0.400(4) & 0.0511(5) \\ 
\hline
\end{tabular}            
\ccaption{}{\label{tab:zqqxs}
$\sigma(\ell^+ \ell^- b \bbar + N~{\rm jets})$
at the Tevatron and at the LHC. Parameters and cuts are given
in eqs.~(\ref{eq:zqq1}-\ref{eq:zqq2}).}
\end{center}
\end{table}}

\subsection{$\mathbf{W+}$~jets}
\label{sec:wjets}
As in the previous cases, we use the notation $W$ as a short hand; what is
actually calculated is the matrix element for a lepton+neutrino final
state. All spin correlations and finite width effects are therefore
accounted for. The quoted cross sections refer to a single lepton
family; in the flavour assignment, the code selects by default an
electron. Different flavours can be selected during the unweighting
phase, covering all possible leptonic decays, as well as inclusive
quark decays (for more details see the Appendix B.4).
The EW parameters are fixed by default using the option {\tt iewopt=3}
(see eq.~(\ref{eq:iew1})).

 The subprocesses considered include all configurations
with up to 2 light-quark pairs; they are listed in
Table~\ref{tab:wjets}, following the notation employed in the
code.
\begin{table}
\begin{center}
\begin{tabular}{ll|ll|ll}
{\tt jproc} & subprocess & {\tt jproc} & subprocess & {\tt jproc} &
subprocess \\ 
1 &  $q\qbar' \to W  $ 
&2 &  $q g \to q' W  $ 
&3 &  $g q \to q' W  $ 
\\
4 &  $gg \to q \qbar' W  $ 
&5 &  $q\qbar' \to W  q'' \qbar'' $ 
&6 &  $qq'' \to W  q' q'' $ 
\\
7 &  $q'' q \to W  q' q'' $ 
&8 &  $q\qbar \to W  q' \qbar'' $ 
&9 &  $q\qbar' \to W  q \qbar $ 
\\
10 &  $\qbar' q\to W  q \qbar $ 
&11 &  $q\qbar \to W  q \qbar' $ 
&12 &  $q\qbar \to W  q' \qbar $ 
\\
13 &  $q q \to W  q q' $ 
&14 &  $q q' \to W  q q $ 
&15 &  $q q' \to W  q' q' $ 
\\
16 &  $q g \to W  q' q''\qbar'' $ 
&17 &  $g q \to W  q' q''\qbar'' $ 
&18 &  $q g \to W  q q \qbar' $ 
\\
19 &  $q g \to W  q' q \qbar $ 
&20 &  $g q \to W  q q \qbar' $ 
&21 &  $g q \to W  q' q \qbar $ 
\\
22 &  $q g \to W q' q' \qbar' $ 
&23 &  $g q \to W q' q' \qbar' $ 
&24 &  $g g \to W q \qbar' q'' \qbar'' $ 
\\
25 &  $g g \to W q \qbar q \qbar' $ 
& &
\end{tabular}
\ccaption{}{\label{tab:wjets} Subprocesses included in the $W+$jets
  code. Additional final-state gluons are not explicitly 
  shown here but are included in the code if the requested light-jet
  multiplicity ($N\le 6$) exceeds the number of indicated final-state partons.
  For example, the subprocess {\tt jproc=1} in the case of 2 jet
  will correspond to the final state  $q\qbar' \to W g g$.
  The details can be found in the subroutine {\tt selflav} of
  the file {\tt wjetlib/wjets.f}.}
\end{center}
\end{table}
For each process, the charge-conjugate ones are always understood.
The above list fully covers all the possible processes with up to 3
light jets in addition to the heavy quarks. In the case of 4 extra
jets, we do not calculate processes with 3 light-quark pairs. Within
the uncertainties of the LO approximation, these can be safely
neglected~\cite{Berends:1991ax}.

As a default, the code generates kinematical configurations defined by
cuts applied to the following variables:
\begin{itemize}
\item $\pt^{\rm jet}$, $\eta^{\rm jet}$, $\Delta R_{jj}$
\item $\pt^{\rm Q}$, $\eta^{\rm Q}$, $\Delta R_{Q\Qbar}$ 
\item $\pt^{\rm \ell}$, $\eta^{\rm \ell}$, $\pt^{\rm \nu}$, $\Delta
  R_{\ell j} \; . $
\end{itemize}
The respective threshold values can be provided by the user at run
time. Additional cuts can be supplied by the user in an appropriate
routine. 
The choice of scale follows the same conventions as for the $WQ\Qbar$
case.

Some benchmark results are given in Table~\ref{tab:wjxs}, obtained
for the following set of cuts and scale choice:
\ba \label{eq:wj1}
&& Q^2 = \mWsq + \ptWsq,
\\
        && \pt^{\rm jet}>20~\gev, \quad \vert \eta_j\vert < 2.5, \quad \Delta
        R_{jj} >0.7 \; .
\label{eq:wj2}
\ea

{\renewcommand{\arraystretch}{1.2}
\begin{table}
\begin{center}
\begin{tabular}{||l|l|l|l|l|l|l|l||}\hline
 & $N = 0$ & $N = 1$ & $N = 2$  & $N = 3$ & $N = 4$ & $N = 5$ & $N = 6$
\\ \hline
LHC (pb) & 18068(4) & 3412(4) & 1130(2) & 342.9(1.4) & 100.6(1.4) & 27.6(4) & 7.14(15)
                 \\  \hline
FNAL (pb)  & 2087.0(6) & 225.8(2) & 37.3(2) &  5.66(6) & 0.745(4) & 0.0864(15) & 0.0086(2)
\\ \hline
\end{tabular}           
\ccaption{}{\label{tab:wjxs} $\sigma(\ell \nu + N~{\rm jets})$
at the Tevatron and 
at the LHC. Parameters and cuts are given
in eqs.~(\ref{eq:wj1}-\ref{eq:wj2}).}
\end{center}
\end{table}}
For processes with up to 4 jets, we verified the numerical agreement
with the results of the \vecbos\ code~\cite{Berends:1991ax}.

\subsection{$\mathbf{W+c+}$~jets}
\label{sec:wcjets}
The processes included in this code are a subset of the ones 
treated in section~\ref{sec:wjets}. Here the final states consisting 
exclusively of one $c$ (or $\bar c$) quark in association with a 
$W$ and additional light jets 
are singled out. Events with $charm$ quark pairs are not included, and
should be generated using the {\tt wqq} processes.
All spin correlations and finite width effects in the fermionic decay
of the $W$ are 
accounted for. The $W$ decay modes can be selected when running 
the code with {\tt imode=2} to produce unweighted events, as discussed
in the appendix B.4. 
The EW parameters are fixed by default using the option {\tt iewopt=3}
(see eq.~(\ref{eq:iew1})).

The subprocesses considered include all configurations
with up to 2 light-quark pairs (where ``light'' includes the charm); 
they are listed in
Table~\ref{tab:wcjets}, following the notation employed in the
code.
\begin{table}
\begin{center}
\begin{tabular}{ll|ll}
{\tt jproc} & subprocess & {\tt jproc} & subprocess \\ 
1 &  $g c' \to W c  $ 
&2 &  $c' g \to W c  $ 
\\
3 &  $gg \to W c \cbar'  $ 
&4 &  $qq'' \to W  q' q'' (q'/q''=c) $ 
\\
5 &  $q'' q \to W  q' q'' (q'/q''=c) $ 
&6 &  $q\qbar \to W  q' \qbar'' (q'/q''=c) $
\\
7 &  $q\qbar \to W  q \qbar' (q/q'=c) $ 
&8 &  $q\qbar \to W  q' \qbar (q/q'=c) $ 
\\
9 &  $q q \to W  q q' (q/q'=c)$ 
&10 &  $c' g \to W  c q''\qbar'' $ 
\\
11 &  $g c' \to W  c q''\qbar'' $ 
&12 &  $c' g \to W  c' c' \cbar $ 
\\
13 &  $c' g \to W  c c' \cpbar $ 
&14 &  $g c' \to W  c' c' \cbar $ 
\\
15 &  $g c' \to W  c c' \cpbar $ 
&16 &  $g g \to W  c \cbar' q'' \qbar'' $ 
\\
17 &  $g g \to W  c' \cpbar c' \cbar $ 
& &
\end{tabular}
\ccaption{}{\label{tab:wcjets} Subprocesses included in the $W+c+$jets
  code. The symbol $c'$ refers to either of the two Cabibbo partners
  of the charm quark, $d$ or $s$.
  Additional final-state gluons are not explicitly 
  shown here but are included in the code if the requested light-jet
  multiplicity ($N\le 5$) exceeds the number of indicated final-state partons.
  For example, the subprocess {\tt jproc=1} in the case of 2 jet
  will correspond to the process  $g c' \to W c g g$.
  The details can be found in the subroutine {\tt selflav} of
  the file {\tt wcjetlib/wcjets.f}.}
\end{center}
\end{table}
For each process, the charge-conjugate ones are always understood.
The above list fully covers all the possible processes with up to 2
light jets in addition to the heavy quarks. In the case of 3 extra
jets, we do not calculate processes with 3 light-quark pairs. As in
the case of $W$+jet production, we expect these to be negligible.

As a default, the code generates kinematical configurations defined by
cuts applied to the following variables:
\begin{itemize}
\item $\pt^{\rm jet}$, $\eta^{\rm jet}$, $\Delta R_{jj}$
\item $\pt^{\rm \ell}$, $\eta^{\rm \ell}$, $\pt^{\rm \nu}$, $\Delta
  R_{\ell j} \; ,$ 
\end{itemize}
where the cuts on $c$-quarks are the same as for light jets. 
The respective threshold values can be provided by the user at run
time. Additional cuts can be supplied by the user in an appropriate
routine. 
The choice of scale follows the same conventions as for the $WQ\Qbar$
case.
Some benchmark results are given in Table~\ref{tab:wcjxs}, obtained
for the following set of cuts and scale choice:
\ba \label{eq:wcj1}
&& Q^2 = \mWsq + \ptWsq,
\\
        && \pt^{\rm jet}>20~\gev, \quad \vert \eta_j\vert < 2.5, \quad \Delta
        R_{jj} >0.7 \; .
\label{eq:wcj2}
\ea

{\renewcommand{\arraystretch}{1.2}
\begin{table}
\begin{center}
\begin{tabular}{||l|l|l|l|l|l|l||}\hline
 & $N = 0$ & $N = 1$ & $N = 2$  & $N = 3$ & $N = 4$ & $N = 5$            \\ \hline
LHC (pb)    & 418.9(8) & 183.7(8) & 55.6(3) & 14.9(1)& 3.65(3)& 0.848(7) \\ \hline
FNAL (fb)   & 8740(20)  & 2390(10)&  360(2)& 42.2(2)& 4.14(2)& 0.353(2)\\ \hline
\end{tabular}           
\ccaption{}{\label{tab:wcjxs} $\sigma(\ell \nu + c +N~{\rm jets})$
at the Tevatron and 
at the LHC. Parameters and cuts are given
in eqs.~(\ref{eq:wcj1}-\ref{eq:wcj2}).}
\end{center}
\end{table}}

\subsection{$\mathbf{Z/\gamma^{*}+}$~jets}
\label{sec:zjets}
We use the notation $Z/\gamma^{*}$ as a short hand; what is actually
calculated is the matrix element for a charged lepton or neutrino pair
final state. All spin correlations and finite width effects are
therefore accounted for. When the final state $\ell^+ \ell^-$ is
selected, the interference between intermediate $Z$ and $\gamma^{*}$ is
also included. The quoted cross sections refer to a single lepton
family; in the flavour assignment, the code selects by default the
$e^+ e^-$ pair. In the case of the final state $\nu \bar\nu$ the
quoted cross sections include the decays to all 3 neutrino flavours,
although we always label the neutrinos as $\nu_e$.
The EW parameters are fixed by default using the option {\tt iewopt=3}
(see eq.~(\ref{eq:iew2})).

All subprocesses with up to 2 light quark pairs are included. 
This means that the cross-sections with up to 3 final-state partons are 
exact. The emission of additional hard gluons can however be
calculated, and the current version of the code works with up to a 6
final-state jets.
The subprocesses considered are listed in Table~\ref{tab:zjets}.

\begin{table}[h]
\begin{center}
\vskip .3cm
\begin{tabular}{ll|ll|ll}
{\tt jproc} & subprocess & {\tt jproc} & subprocess & {\tt jproc} &
subprocess \\ 
1  &  $\qu \qubar \to   Z$ 
&2 &  $\qd \qdbar \to   Z$ 
&3 &  $ g \qu \to \qu Z $ 
\\
4  &  $g \qd  \to \qd Z$ 
&5 &  $\qu g  \to \qu Z $ 
&6 &  $\qd g  \to \qd Z $ 
\\
7  &  $g g  \to \qu \qubar Z$ 
&8 &  $g g  \to \qd \qdbar Z$ 
&9 &  $\qu \qubar \to \qu \qubar Z$ 
\\
10  &  $\qd \qdbar \to \qd \qdbar Z  $ 
&11 &  $ \qu \qu   \to \qu \qu Z$ 
&12 &  $ \qd \qd   \to \qd \qd Z$ 
\\               
13 &   $\qu \qubar \to \qu' \qubar' Z $ 
&14 &  $\qd \qdbar \to \qd' \qdbar' Z  $ 
&15 &  $\qu \qu' \to \qu \qu' Z$ 
\\
16  &  $\qu \qubar'  \to \qu \qubar' Z $ 
&17 &  $\qd \qd'     \to \qd \qd' Z $ 
&18 &  $ \qd \qdbar' \to \qd \qdbar' Z $ 
\\
19 &   $\qu \qubar \to \qd \qdbar Z $ 
&20 &  $\qd \qdbar \to \qu \qubar Z$ 
&21 &  $\qu \qd    \to \qu \qd Z$ 
\\               
22 & $\qd \qu    \to \qu \qd Z $ 
& 23 & $\qu \qdbar \to \qu \qdbar Z$ 
& 24 & $\qd \qubar \to \qd \qubar Z$
\\ 
25 & $g \qu     \to \qu \qu  \qubar Z   $
& 26 & $\qu g   \to \qu \qu  \qubar Z   $
& 27 & $g \qu   \to \qu \qu' \qubar' Z   $
\\
28 & $\qu g     \to \qu \qu' \qubar' Z   $
& 29 & $g \qu   \to \qu \qd  \qdbar Z   $
& 30 & $\qu g   \to \qu \qd  \qdbar Z   $
\\
31 & $g \qd     \to \qd \qd  \qdbar Z   $
& 32 & $\qd g   \to \qd \qd  \qdbar Z  $
& 33 & $g \qd   \to \qd \qd' \qdbar' Z   $
\\
34 & $\qd g     \to \qd \qd' \qdbar' Z   $
& 35 & $g \qd   \to \qd \qu  \qubar Z   $
& 36 & $\qd g   \to \qd \qu  \qubar Z   $
\end{tabular}
\ccaption{}{\label{tab:zjets} Subprocesses included in the
  $Z/\gamma^{*}+$jets code.  It is always understood that quarks $u$ and
  $d$ represent generic quarks of type up or down. The $Z$ in the
  table stands for a neutral $\ell^+ \ell^-$ ($\nu \bar\nu$) lepton
  pair.  The complex conjugate processes are also understood.
  Additional final-state gluons are not explicitly shown here but are
  included in the code if the requested jet multiplicity ($N\le 6$)
  exceeds the number of indicated final-state partons.  For example,
  the subprocess {\tt jproc=1} in the case of 2  jets will
  correspond to the final state $u\ubar \to Z  g g$.  The
  details can be found in the subroutine {\tt selflav} of the file
  {\tt zjetlib/zjets.f}.}
\end{center}
\end{table}

As a default, the code generates kinematical configurations defined by
cuts applied to the following variables:
\begin{itemize}
\item $\pt^{\rm jet}$, $\eta^{\rm jet}$, $\Delta R_{jj}$
\item $\pt^{\ell}$, $\eta^{\ell}$, $\Delta R_{\ell j}$ for $\ell^+ \ell^-$ 
final states
\item missing $E_T$ for $\nu \bar\nu$ final states.  
\end{itemize}
Additional cuts can be supplied by the user in an appropriate routine.
The choice of scale follows the same conventions as for the $ZQ\Qbar$
case.

Some benchmark results are given in Table~\ref{tab:zjxs}, obtained
for the following set of cuts and scale choice:
\ba \label{eq:zj1}
&& Q^2 = \mZsq + \ptZsq,
\\
        && \pt^{\rm jet}>20~\gev, \quad \vert \eta_j\vert < 2.5, \quad \Delta
        R_{jj} >0.7
\\
        && 80~\gev \leq m_{ll} \leq 100~\gev \; .
\label{eq:zj2}
\ea

{\renewcommand{\arraystretch}{1.2}
\begin{table}
\begin{center}
\begin{tabular}{||l|l|l|l|l|l|l|l||}\hline
 & $N = 0$  & 
$N = 1$ & $N = 2$ & $N = 3$ & $N = 4$ & $N = 5$ & $N = 6$ \\ 
\hline
LHC (pb)  &  1526(1) &  320.9(5) & 104.6(2) & 31.6(2) & 9.4(2) & 
2.51(4) & 0.65(2) \\ 
\hline
FNAL (pb)  & 179.4(2) & 21.44(2) & 3.36(1) & 0.489(2) & 
0.0630(3) & 0.00700(4)& 0.000690(6) \\ 
\hline
\end{tabular}            
\ccaption{}{\label{tab:zjxs} $\sigma(\ell^+ \ell^- + N~{\rm jets})$
at the Tevatron and at LHC. Parameters and cuts are given
in eqs.~(\ref{eq:wj1}-\ref{eq:wj2}).}
\end{center}
\end{table}}

\subsection{$\mathbf{ nW+mZ+lH+k}$ jets}
\label{sec:nw}
The code computes all processes where massive EW gauge bosons and/or Higgs
particles are produced on shell plus (up to 3) additional light jets.
The limitations in the number of final state particles is therefore
the following: $ k \le 3$ and $n+m+l+k \le 8$. It is intended that at
least one EW gauge boson or Higgs particle appear in the final state 
 ($n+m+l>0$). All contributions from
QCD and EW processes are included. In particular, all processes of
the gauge-boson-fusion type are present when the number of final state
quarks is at least 2 (e.g. $qq\to qqH$).
All gauge bosons are decayed to fermion pairs, taking therefore into account 
all spin correlations among the decay products by means of exact 
matrix elements. More information on gauge-invariance issues in
presence of decays are discussed below, and details on the choice of
the decay modes can 
be found in Appendix B.4. The Higgs boson decays will be soon available. 

In the case of $m=l=0$ ($n=l=0$) the code reproduces the results of
the programs described in Sections~\ref{sec:wjets}
and~\ref{sec:zjets}, up to the following effects: final-state
finite-width corrections are absent here since the gauge bosons are
kept on shell; branching ratios depend on the selected decay mode for
the $Z$, while they are kept to 1 in the case of the $W$ bosons, whose
final states are selected during the unweighting phase (see Appendix
B4); the $\gamma^* \to \ell^+\ell^- $ contributions are only present
in the code of Section~\ref{sec:zjets}; EW production of jets ($W\to
jj$, $Z/\gamma^* \to jj$, plus EW boson exchanges between quarks) are
included here.

The EW parameters are fixed by default using the option {\tt iewopt=3}
(see eq.~(\ref{eq:iew3})). In this way we are guaranteed that all gauge
boson masses are consistent with the measured values. The coupling
strengths extracted from the fixed inputs, however, will differ by few
percents from their best values. The user can select the option 
 {\tt iewopt=0}
(see eq.~(\ref{eq:iew0})), where the couplings are matched to the
radiatively corrected best values, at the price of working with gauge
boson masses which differ by few percents from the measured
masses. Both schemes are equally consistent at the LO. We verified
that cross-sections obtained using the two schemes differ from each
other at the level of few percent, a negligible effect compared to
the large uncertainties related to the choice of factorization and
renormalization scales.

The subprocesses considered are listed in Table~\ref{tab:WZH}, where a
distinction is made according to the number of final state $W$'s.  If
$n$ is odd, the whole Cabibbo structure of the matrix element is
correctly taken into account. If $n$ is even, we work in the $\cos
\theta_c= 1$ approximation. The reason of this choice is that, in the
latter case, different Cabibbo structures may interfere at the
amplitude level. The quantity ${\cal F}^{\pm}$ in Table~\ref{tab:WZH}
stands for $nW+mZ+lH$ with $n$ odd, while ${\cal F}^{0}$ stands for
$nW+mZ+lH$ when $n$ is even.

\begin{table}
\begin{center}
\vskip .3cm
{\bf $\mathbf{n}$ odd}
\vskip .3cm
\begin{tabular}{ll|ll|ll}
{\tt jproc} & subprocess & {\tt jproc} & subprocess & {\tt jproc} &
subprocess \\ 
1  &  $q\qbar' \to {\cal F}^{\pm} $ 
&2 &  $g q \to q' {\cal F}^{\pm} $ 
&3 &  $q g \to q' {\cal F}^{\pm} $ 
\\
4 &   $gg \to q \qbar' {\cal F}^{\pm} $ 
&5 &  $q\qbar' \to {\cal F}^{\pm} q'' \qbar'' $ 
&6 &  $qq'' \to {\cal F}^{\pm} q' q'' $ 
\\
7 &  $q'' q \to {\cal F}^{\pm} q' q'' $ 
&8 &  $q''\qbar'' \to {\cal F}^{\pm} q \qbar' $ 
&9 &  $q\qbar' \to {\cal F}^{\pm} q \qbar $ 
\\
10 &  $\qbar' q\to {\cal F}^{\pm} q \qbar $ 
&11 &  $q\qbar \to {\cal F}^{\pm} q \qbar' $ 
&12 &  $q\qbar \to {\cal F}^{\pm} q' \qbar $ 
\\
 13 &  $q q  \to {\cal F}^{\pm} q q' $ 
&14 &  $q q' \to {\cal F}^{\pm} q q $ 
&15 &  $q q' \to {\cal F}^{\pm} q' q' $ 
\\
16  &  $q g \to {\cal F}^{\pm} q' q''\qbar'' $ 
&17 &  $g q \to {\cal F}^{\pm} q' q''\qbar'' $ 
&18 &  $q g \to {\cal F}^{\pm} q q \qbar' $ 
\\
19 &  $q g \to  {\cal F}^{\pm} q' q \qbar $ 
&20 &  $g q \to {\cal F}^{\pm} q q \qbar' $ 
&21 &  $g q \to {\cal F}^{\pm} q' q \qbar $ 
\end{tabular}
\vskip .3cm
{\bf $\mathbf{n}$ even}
\vskip .3cm
\begin{tabular}{ll|ll|ll}
{\tt jproc} & subprocess & {\tt jproc} & subprocess & {\tt jproc} &
subprocess \\ 
1  &  $q\qbar \to   {\cal F}^0$ 
&2 &  $g q    \to q {\cal F}^0$ 
&3 &  $q g    \to q {\cal F}^0$ 
\\
4  &  $gg     \to q \qbar {\cal F}^0$ 
&5 &  $qq     \to q q     {\cal F}^0$ 
&6 &  $qq     \to q' q'   {\cal F}^0$ 
\\
7  &  $q q'  \to  q q'      {\cal F}^0$ 
&8 &  $q q'' \to  q q''     {\cal F}^0$ 
&9 &  $q q'  \to  q'' q'''  {\cal F}^0$ 
\\
10  &  $q \qbar \to   q \qbar     {\cal F}^0$ 
&11 &  $q \qbar \to   q' \qbar'   {\cal F}^0$ 
&12 &  $q \qbar \to   q'' \qbar'' {\cal F}^0$ 
\\
13 &   $q \qbar'  \to   q \qbar' {\rm~or~} \qbar q' {\cal F}^0$ 
&14 &  $q \qbar'' \to   q \qbar''                  {\cal F}^0$ 
&15 &  $q \qbar'  \to   q'' \qbar'''               {\cal F}^0$ 
\\
16  &  $q  \qbar'' \to q' \qbar''' {\cal F}^0$ 
&17 &  $g q  \to q q \qbar {\cal F}^0 $ 
&18 &  $q g  \to q q \qbar {\cal F}^0 $ 
\\
19 &   $g q  \to q q' \qbar' {\cal F}^0$ 
&20 &  $q g  \to q q' \qbar' {\cal F}^0$ 
&21 &  $g q  \to q q'' \qbar''{\cal F}^0 $ 
\\
  22 & $q g  \to q q'' \qbar'' {\cal F}^0$ 
& 23 & $g q  \to q' q' \qbar   {\cal F}^0$ 
& 24 & $q g  \to q' q' \qbar   {\cal F}^0$ 
\\
  25 & $g q  \to q' q'' \qbar''' {\cal F}^0$ 
& 26 & $q g  \to q' q'' \qbar''' {\cal F}^0$ 
& &
\end{tabular}
\ccaption{}{\label{tab:WZH} Subprocesses included in the $nW+mZ+lH+$jets
  code, for the cases n=odd ($nW+mZ+lH={\cal F}^{\pm}$)
and n=even ($nW+mZ+lH={\cal F}^0$).
It is always understood that quarks $q$ and $q'$ belong to the same
iso-doublet, while $q$ and $q''$ belong to different iso-doublets.
  Additional final-state
  gluons are not explicitly indicated but are included in the
  code. 
  The details can be found in the subroutine {\tt selflav} of
  the file {\tt vbjetslib/vbjets.f}.}
\end{center}
\end{table}

As a default, the code generates kinematical configurations defined by
cuts applied to the following variables:
\begin{itemize}
\item $\pt^{\rm jet}$, $\eta^{\rm jet}$, $\Delta R_{jj} \; .$
\end{itemize}
Additional cuts can be supplied by the user in an appropriate
routine. Cuts on  the leptons or on  the Higgs decay products
should be imposed in the analysis routine, using the momentum and
flavour information as specified in Appendix B.4.

In the code initialization phase, the user can select among 3
continuous choices for the parametrization of the factorization and
renormalization scale $Q$: a real input parameter ({\tt qfac}) allows
to vary the overall scale of $Q$, $Q={\tt qfac}\times Q_0$, while the
preferred functional form for $Q_0$ is selected through an integer
input parameter ({\tt iqopt}=0,1,2).  In more detail:
{\renewcommand{\arraystretch}{1.2}
\begin{center}
\begin{tabular}{l||l|l|l}
{\tt iqopt} & 0 & 1 & 2 \\ \hline
$Q_0$ & $\langle M_B \rangle$ & $\sum M_B$ & $\sqrt{\hat{s}}$
\end{tabular}
\end{center}
}
where $\langle M_B \rangle$ and $\sum M_B$ are the average and sum of
the EW boson masses (vectors as well as Higgses), respectively.

Some benchmark results are given in
Tables~\ref{tab:wwxs}-\ref{tab:whxs}, using the scale $Q = \sum M_B$,
with the sum extended over all bosons in the final state.

In order to deal, in a gauge invariant way, with the problem of the 
unstable bosons appearing in the intermediate states, we computed 
the matrix element by using zero-width propagators and by
cutting away events around the mass of the unstable particle 
$M$ in such a way to keep the area of a Breit-Wigner distribution.
In other words, for a particle of mass 
$M$ and width $\Gamma$, our effective cut $s_0$ 
is determined by the equation  
\bqa
\int_{-\infty}^{M^2-s_0} ds \frac{1}{(s-M^2)^2} = 
\int_{-\infty}^{M^2} ds \frac{1}{(s-M^2)^2+\Gamma^2 M^2} \; ,
\eqa
giving $s_0= \frac{2\Gamma M}{\pi}$. In practice the 
program discards events for which $|s-M^2| < {\tt winsize}\times
\Gamma  M$, with the adjustable parameter ${\tt winsize}$ set, by
default, to the value ${\tt winsize}= \frac{2}{\pi}$.
The described procedure gives sensible results 
when the portion of resonance cut away is of the order of a few GeV. 
Otherwise, holes start becoming visible in the distributions.
This is safe for $Z$ and $W$ vector bosons. But problems arise
for very heavy Higgses. For this reason we put a protection in the code
to inhibit the calculation when $\Gamma_H > 10$ GeV.

We checked that our algorithm reproduces, within few percents, 
the results obtained by using the following  approach:  we set
$\Gamma= 0$ in the matrix element and multiply
the final result by the factor 
$$ \frac{1}{1+\left(\frac{M\Gamma}{s-M^2}\right)^2}\; .$$


As a consistency check of our calculations for large multiplicities of
gauge bosons, we verified that the production rates for multiple gauge
bosons when the Higgs mass is above the threshold for diboson decay
are well approximated by the incoherent sum of the processes mediated
by an on-shell Higgs, plus those where the Higgs contribution 
is suppressed in the
intermediate states. In the tables we show, this is for example seen
in the comparison of the $WWjj$ vs $Hjj$ or $VVV$ vs $HV$
 rates at \mH=200~GeV.
In the case of 3 gauge boson production, we verified the agreement
with the results shown in ref.~\cite{Haywood:1999qg}, 
up to the $WZZ$ channel, whose
rate is erroneously reported in the tables of 
Section 5.33\footnote{A. Ghinkulov, private communication.}. In the
case of four boson production, a previous calculation has been
documented in \cite{Barger:1989cp}. A comparison of the results of our
code with the numbers presented in that paper shows however some
discrepancies.

{\renewcommand{\arraystretch}{1.2}
\begin{table}
\begin{center}
\begin{tabular}{||l|l|l|l||}\hline
& $WW$ & $WZ$ & $ZZ$ 
                 \\  \hline
LHC (pb)   & 74.8(5) & 28.1(2) & 10.85(6) 
\\ \hline
FNAL (pb)  & 8.51(5) & 2.45(1) & 1.027(5)
\\ \hline
\end{tabular}            
\ccaption{}{\label{tab:wwxs} Diboson production 
at the Tevatron and at the LHC. }
\end{center}
\end{table}}

{\renewcommand{\arraystretch}{1.2}
\begin{table}
\begin{center}
\begin{tabular}{||l|l|l|l|l||}\hline
 $WW+$jets & $jj$, central & 
            $jjj$, central 
& $jj$, fwd 
& $jjj$,fwd 
                 \\  \hline
LHC, \mH=120 (pb)   & 18.6(2) &8.16(9) &0.256(4) & 0.365(9)
\\ \hline	     	    
LHC, \mH=200 (pb)   & 18.9(2) &8.3(1)  &0.546(24)& 0.389(9)
\\ \hline	     	    
FNAL, \mH=120 (fb)  & 336(1)  &49.1(2) &0.201(1) & 0.0789(3) \\ \hline
FNAL, \mH=200 (fb)  & 364(2)  &54.9(8) &0.415(4) & 0.096(2) \\ \hline
\end{tabular}            
\ccaption{}{\label{tab:wwjxs} Associated production of $WW$ and jets,
at the Tevatron and at the LHC. In all cases, $\et_j>20~\gev$ and
$\Delta R_{jj}>0.7$. The {\em central} configurations correspond to
all jets with $\vert \eta_j \vert<2.5$. The {\em fwd} configurations
have two jets in opposite rapidity hemispheres 
with $2.5 < \vert \eta_j \vert<5$, plus a central jet in the $jjj$ case.}
\end{center}
\end{table}}

{\renewcommand{\arraystretch}{1.2}
\begin{table}       
\begin{center}
\begin{tabular}{||l|l|l|l|l||}\hline
 $H+$jets & $jj$, central & 
           $jjj$, central
& $jj$, fwd
& $jjj$,fwd
                 \\  \hline
LHC, \mH=120 (pb)   &1.27(1)  &0.458(6)& 0.504(3) &0.089(1) 
\\ \hline	       	       	  
LHC, \mH=140 (pb)   &0.96(1)  &0.320(4) & 0.458(2) &0.078(2) 
\\ \hline	       	       	  
LHC, \mH=200 (pb)   &0.459(4) &0.132(3) & 0.346(2) &0.0528(6) 
\\ \hline	       	       	  
FNAL, \mH=120 (fb)  & 121(1)  &26.7(2) & 0.769(2) &0.0730(5) 
\\ \hline	       	       	  
FNAL, \mH=140 (fb)  & 75.0(4) &16.0(1) & 0.573(1) &0.0513(3) 
\\ \hline	       	       	  
FNAL, \mH=200 (fb)  &21.8(2)  &4.28(3) & 0.246(1) &0.0193(1) 
\\ \hline
\end{tabular}            
\ccaption{}{\label{tab:hjxs} Associated production of $H$ and jets,
at the Tevatron and at the LHC. In all cases, $\et_j>20~\gev$ and
$\Delta R_{jj}>0.7$. The {\em central} configurations correspond to
all jets with $\vert \eta_j \vert<2.5$. The {\em fwd} configurations
have two jets in opposite rapidity hemispheres 
with $2.5 < \vert \eta_j \vert<5$, plus a central jet in the $jjj$ case.}
\end{center}
\end{table}}

{\renewcommand{\arraystretch}{1.2}
\begin{table}
\begin{center}
\begin{tabular}{||l|l|l|l|l||}\hline
        & $WWW$ & $WWZ$ & $WZZ$ & $ZZZ$ 
   \\  \hline
LHC, \mH=120  & 130(1)&98(2)&31.0(4) &10.9(1)
   \\  \hline 	       	      
LHC, \mH=200  & 305(5)&199(4)&95(1) & 45.2(5)
   \\  \hline	       	      
FNAL, \mH=120 & 6.39(2)&5.13(4)&1.21(1) &0.519(2)
   \\  \hline	       	      
FNAL, \mH=200 & 18.1(2)&13.5(2)&5.47(6) &3.36(3)
  \\  \hline
\end{tabular}            
\ccaption{}{\label{tab:wwwxs} Triboson production at the 
Tevatron and 
at the LHC (fb).}
\end{center}
\end{table}}

{\renewcommand{\arraystretch}{1.2}
\begin{table}
\begin{center}
\begin{tabular}{||l|l|l|l|l|l||}\hline
        & $WWWW$ & $WWWZ$ & $WWZZ$ & $WWWWW$ & $WWWWZ$ 
                 \\  \hline
LHC, \mH=120 & 0.606(6) & 0.72(1) & 0.48(1) & $5.8(1)\cdot 10^{-3}$
                                            & $10.8(3)\cdot 10^{-3}$
                 \\  \hline
\end{tabular}            
\ccaption{}{\label{tab:4wxs} 4- and 5-boson production at the 
LHC (fb). }
\end{center}
\end{table}}

{\renewcommand{\arraystretch}{1.2}
\begin{table}
\begin{center}
\begin{tabular}{||l|l|l|l||l|l|l|l||}\hline
$WH$ & \mH=120 & \mH=140 & \mH=200 &
$ZH$ & \mH=120 & \mH=140 & \mH=200 
\\ \hline
LHC (pb)   & 1.364(8) & 0.833(4) & 0.251(2) & & 0.727(4) & 0.449(2) & 0.137(1)
\\ \hline
FNAL (fb)  & 122.4(6)& 70.0(2) & 16.55(4) & & 75.0(3) & 44.2(2) & 11.24(3)
\\ \hline
\end{tabular}            
\ccaption{}{\label{tab:whxs} H+boson production 
at the Tevatron and LHC.}
\end{center}
\end{table}}

\subsection{$\mathbf{ Q\Qbar+}$ jets}
\label{sec:2Q}
The subprocesses considered include all configurations with up to two
light quark pairs. They are listed in Table~\ref{tab:2Q}, following 
the notations employed in the code. The list covers all possible 
processes involving up to 4 light-parton jets with 2 light quark pairs.
Subprocesses involving 3 light-quark pairs, which would only appear in
the case of 4 jets in addition to the heavy quarks, are not included,
as they are expected to contribute a negligible rate.

As a default, the code generates kinematical configurations defined by
cuts applied to the following variables (the cuts related to the heavy
quarks are only applied in the case of $b$, while top quarks are
always generated without cuts):
\begin{itemize}
\item $\pt^{\rm jet}$, $\eta^{\rm jet}$, $\Delta R_{jj},\Delta R_{jb}$
\item $\pt^{ b}$, $\eta^{ b}$, $\Delta R_{b\bbar} \; .$
\end{itemize}
The respective threshold values can be provided by the user at run
time. Additional cuts can be supplied by the user in an appropriate
routine. In the case $Q=t$ the decay of $t \bar t$ pairs in six fermions 
(in the on-shell approximation) is enforced.
With this option the decay of the top quarks takes into account 
all spin correlations among the decay products by means of exact 
matrix elements. More information on the selection of decay products can 
be found in Appendix B.4.

In the code initialization phase, the user can select 
between 2 choices for the parameterization of the 
factorization and renormalization scale $Q$. A real input
parameter ({\tt qfac}) allows to vary the overall scale of $Q$,
$Q={\tt qfac}\times Q_0$, while the preferred functional form for
$Q_0$ is selected through an integer input parameter ({\tt
iqopt}=0,1).  In more detail:
{\renewcommand{\arraystretch}{1.2}
\begin{center}
\begin{tabular}{l||l|l}
{\tt iqopt} & 0 & 1  \\  \hline
$Q_0[tt]^2$ & $m_t^2$ & $m_t^2+\langle \ptsq \rangle$ \\
$Q_0[bb]^2$ & $\hat{s}$ & $\langle \ptsq \rangle$ \\
\end{tabular}
\end{center}}
where $\langle \pt \rangle $ is the average $\pt$ of light partons, 
if $Q=t$, while is the average transverse momentum 
of all final state jets in the case of $Q=b$.

\begin{table}
\begin{center}
\begin{tabular}{ll|ll|ll}
{\tt jproc} & subprocess & {\tt jproc} & subprocess & {\tt jproc} &
subprocess \\ 
1 &  $g g \to  Q\Qbar$ 
&2 &  $q \qbar \to Q\Qbar$ 
&3 &  $g q \to  Q\Qbar q$ 
\\
4 &  $ q g \to Q\Qbar q$ 
&5 &  $g g  \to  Q \Qbar q \qbar$ 
&6 &  $q \qbar  \to Q \Qbar q \qbar$ 
\\
7 &  $q \qbar \to Q \Qbar q' \qbar'$ 
&8 &  $q q' \to Q \Qbar q q'$ 
&9 &  $q\qbar' \to  Q \Qbar q \qbar'$ 
\\
10 &  $q q\to Q \Qbar q q$ 
&11 &  $g q \to Q \Qbar q q'\qbar'$ 
&12 &  $q g  \to  Q \Qbar q q' \qbar'$ 
\\
13 &  $g q \to  Q \Qbar q q \qbar$ 
&14 &  $q g \to Q \Qbar q q \qbar $ 
&15 &  $g g \to  Q \Qbar q \qbar q \qbar$ 
\\
16 &  $g g \to  Q \Qbar q \qbar q' \qbar'$ 
& &  
& &
\end{tabular}
\ccaption{}{\label{tab:2Q} Subprocesses included in the $Q\Qbar+$jets
  code. Additional final-state gluons are not explicitly 
  shown here but are included in the code if the requested light-jet
  multiplicity ($N\le 6$) exceeds the number of indicated final-state partons.
  For example, the subprocess {\tt jproc=1},
  in the case of 2 extra jets, will correspond to the final state  
  $gg\to Q\Qbar g g$.
  The details can be found in the subroutine {\tt selflav} of
  the file {\tt 2Qlib/2Q.f}. For each process, the charge-conjugates 
  ones are always understood.}
\end{center}
\end{table}

Some benchmark results are given in Table~\ref{tab:QQxs}, obtained
for the following set of cuts and scale choice:
\ba \label{eq:2Qcut1}
t\tbar: && \mt=175~\gev, Q^2=\mtsq
\\
b\bbar: && \mb=4.75~\gev, Q^2=(\ptbsq+\ptbbsq+\sum \ptsq_j)/(2+N)
\\
        && \pt^{\rm jet}>20~\gev, \quad \vert \eta_j\vert < 2.5, \quad \Delta
        R_{jj} >0.7
\\
        && \pt^{b}>20~\gev, \quad \vert \eta_b \vert < 2.5, \quad \Delta
        R_{b\bbar} >0.7, \Delta R_{bj} >0.7 \; .
\label{eq:2Qcut2}
\ea
In the case of 0 and 1 jet, we find agreement with the results obtained
using the \oacube\ code of ref.~\cite{Mangano:jk}.
{\renewcommand{\arraystretch}{1.2}
\begin{table}
\begin{center}
\begin{tabular}{||l|l|l|l|l|l|l|l||}\hline
$Q \Qbar + N~{\rm jets}$  & $N= 0$  & 
$N = 1$ & $N = 2$ & $N = 3$ & $N=4$ & $N=5$ & $N=6$ \\ 
\hline
$Q=t$, LHC (pb)  & 530.0(8) & 462.6(6) & 255(1) & 111.5(5) & 42.4(4)  
& 14.07(16) & 4.36(8) \\ 
\hline
$Q=t$, FNAL (fb) &  6,364(8) &  1,592(3) & 282(1) & 40.6(3) & 4.83(4) & 
0.483(6) & 0.0419(9) \\ 
\hline
$Q=b$, LHC (nb)   &  1,533(4)
& 422(1) & 130.2(6)  & 30.9(4) & 7.5(4) 
& 1.53(6) & 0.337(9) \\ 
\hline
$Q=b$, FNAL (pb)   & 72,1(1) &  12,15(2) & 2,51(1) & 365(7) & 47.3(9) 
& 5.6(2)& 0.58(2)\\ 
\hline
\end{tabular}            
\ccaption{}{\label{tab:QQxs} $\sigma(Q \Qbar + N~{\rm jets})$
at the Tevatron and 
at the LHC, with parameters and cuts given in
eqs.~(\ref{eq:2Qcut1}-\ref{eq:2Qcut2}).}
\end{center}
\end{table}}

\subsection{$\mathbf{ Q\Qbar Q'\Qbar'+}$ jets}
\label{sec:4Q}
The list of available processes is given in Table~\ref{tab:4Q}. This
covers all possible processes up to 2 light jets and at most 1 light
quark pair.  The cases with one additional heavy quark pair (e.g.
$t\tbar b\bbar b\bbar$) are also included. Subprocesses with two
light-quark pairs give a negligible contribution to the case of 2
extra jets, and are not calculated.  For the parameterization of the
factorization and renormalization scale the user has two choices, {\tt
  iqopt}=0,1, as described in the following Table:
{\renewcommand{\arraystretch}{1.2}
\begin{center}
\begin{tabular}{l||l|l}
{\tt iqopt} & 0 & 1  \\ \hline
$Q_0[bb]^2$ & $\hat{s}$ & $\langle \ptsq  \rangle $ \\
$Q_0[bt]^2$ & $\hat{s}$ & $m_t^2$ \\
$Q_0[tt]^2$ & $\hat{s}$ & $m_t^2$ \\
\end{tabular}
\end{center}}
where $\langle \ptsq  \rangle $ is the average $\ptsq $ of light partons and
$b$'s. 

Again, as a default, the code generates kinematical configurations defined by
cuts applied to the following variables (the cuts related to the heavy
quarks are only applied in the case of $b$, while top quarks are
always generated without cuts):
\begin{itemize}
\item $\pt^{\rm jet}$, $\eta^{\rm jet}$, $\Delta R_{jj}, \Delta R_{jb}$
\item $\pt^{ b}$, $\eta^{ b}$, $\Delta R_{b\bbar} \; .$
\end{itemize}
The respective threshold values can be provided by the user at run
time. Additional cuts can be supplied by the user in an appropriate
routine. 

\begin{table}
\begin{center}
\begin{tabular}{ll|ll|ll}
{\tt jproc} & subprocess & {\tt jproc} & subprocess & {\tt jproc} &
subprocess \\ 
1 &  $g g \to  Q\Qbar Q'\Qbar'$ 
&2 &  $q \qbar \to Q\Qbar Q'\Qbar' $ 
&3 &  $g q \to  Q\Qbar Q'\Qbar' q$ 
\\
4 &  $ q g \to Q\Qbar Q'\Qbar' q$ 
&5 &  $g g  \to  Q \Qbar Q'\Qbar' q \qbar$ 
&6 &  $g g  \to Q \Qbar Q'\Qbar' b {\bar{b}}$ 
\end{tabular}
\ccaption{}{\label{tab:4Q} Subprocesses included in the $Q\Qbar Q'\Qbar'+$
  jets code. Additional final-state gluons are not explicitly 
  shown here but are included in the code if the requested light-jet
  multiplicity ($N\le 4$) exceeds the number of indicated final-state partons.
 For example, the subprocess {\tt jproc=1}, in the case of 2 extra 
  jets, will correspond to the final state  $gg\to Q\Qbar Q'\Qbar' g g$.
  The details can be found in the subroutine {\tt selflav} of
  the file {\tt 4Qlib/4Q.f}. For each process, the charge-conjugates 
  ones are always understood. }
\end{center}
\end{table}
Some benchmark results are given in Table~\ref{tab:4Qxs}, obtained
for the following set of cuts and scale choice:
\mt=175~\gev, \mb=4.75~\gev, 
\ba \label{eq:4Qcut1}
t\tbar t\tbar {\rm ~and~} t\tbar b\bbar : && Q^2=\mtsq
\\
b\bbar b\bbar: && Q^2=(\ptbsq+\ptbbsq+\sum \ptsq_j)/(2+N)
\\
        && \pt^{\rm jet}>20~\gev, \quad \vert \eta_j\vert < 2.5, \quad \Delta
        R_{jj} >0.7
\\
        && \pt^{b}>20~\gev, \quad \vert \eta_b \vert < 2.5, \quad \Delta
        R_{b\bbar} >0.7, \Delta R_{bj} >0.7 \; .
\label{eq:4Qcut2}
\ea

{\renewcommand{\arraystretch}{1.2}
\begin{table}
\begin{center}
\begin{tabular}{||l|l|l|l|l|l||}\hline
$Q \Qbar Q' \Qbar' + N~{\rm jets}$  & $N = 0$  & 
$N = 1$ & $N = 2$ & $N = 3$ & $N = 4$\\ 
\hline
$t \tbar t \tbar$, LHC (fb) &  12.73(8) & 17.4(2) & 13.5(1) & 7.55(6) 
& 3.48(5)\\ 
\hline
$t \tbar b \bbar$, LHC (pb)   &  1.35(1) & 1.47(2) &  
0.94(2) & 0.457(8) & 0.189(4)\\ 
\hline
$t \tbar b \bbar$, FNAL (fb)  & 3.44(3) & 0.95(1) & 0.154(1) & 
0.0187(2) & 0.00187(5) \\ 
\hline
$b \bbar b \bbar$, LHC (pb) & 477(2)  & 259(5) & 95(1) & 28.6(6) &
25.0(3) \\ 
\hline
$b \bbar b \bbar$, FNAL (pb) & 6.64(5) & 2.25(3) & 0.470(5) & 0.076(1) 
& 0.0025(5) \\ 
\hline
\end{tabular}            
\ccaption{}{\label{tab:4Qxs} $\sigma(Q \Qbar Q' \Qbar' + N~{\rm jets})$
at the Tevatron and 
at the LHC, with parameters and cuts given in
eqs.~(\ref{eq:4Qcut1}-\ref{eq:4Qcut2}).}
\end{center}
\end{table}}

\subsection{$\mathbf{Q \Qbar H +}$ jets}
\label{sec:QQH}
The list of processes is given in 
Table~\ref{tab:QQH}. The Higgs is produced only via Yukawa couplings
to the heavy quarks, no other EW process is included.
All cases with up to 2 light-quark pairs are
included, covering in full the possible final states with up to 3
jets in addition to the $QQH$ system. The code will otherwise deal
with up to 4 extra jets.
\begin{table}
\begin{center}
\begin{tabular}{ll|ll|ll}
{\tt jproc} & subprocess & {\tt jproc} & subprocess & {\tt jproc} & 
subprocess \\  
1  &  $g g \to  Q\Qbar H$   &
2 &  $q \qbar \to Q\Qbar H$ &
3 &  $g q \to Q\Qbar q H$ \\
4 &  $q g \to Q\Qbar q H$ &
5 &  $g g \to Q\Qbar q \qbar H$ &
6 &  $g g \to Q\Qbar b \bbar H$ \\
7 &  $q q \to Q\Qbar q q H$ &
8 &  $q q' \to Q\Qbar q q' H$ &
9 &  $q \qbar \to Q\Qbar q \qbar H$ \\
10 &  $q \qbar' \to Q\Qbar q \qbar' H$ &
11 &  $q \qbar \to Q\Qbar q' \qbar' H$ &
12 &  $q \qbar \to Q\Qbar b \bbar H$ \\
13 &  $g q \to Q\Qbar q \qbar q H$ &
14 &  $q g \to Q\Qbar q \qbar q H$ &
15 &  $g q \to Q\Qbar q' \qbar' q H$  \\
16 &  $q g \to Q\Qbar q' \qbar' q H$ &
17 &  $g q \to Q\Qbar b \bbar q H$ &
18 &  $q g \to Q\Qbar b \bbar q H$  \\
\end{tabular}
\ccaption{}{\label{tab:QQH} Subprocesses included in the $Q\Qbar H$ code. 
  The details can be found in the subroutine {\tt selflav} of
  the file {\tt QQhlib/QQh.f}. For each process, the charge-conjugate
  subprocesses are always understood.}
\end{center}
\end{table}

As a default, the code generates kinematical configurations defined by
cuts applied to the following variables:
\begin{itemize}
\item $\pt^{\rm jet}$, $\eta^{\rm jet}$, $\Delta R_{jj}, \Delta R_{jb}$
\item $\pt^{ b}$, $\eta^{ b}$, $\Delta R_{b\bar{b}} \; .$
\end{itemize}
The respective threshold values can be provided by the user at run
time. Additional cuts can be supplied by the user in an appropriate
routine. In the case $Q=t$ the decay of $t \bar t$ pairs in six fermions 
(in the on-shell approximation). 
With this option the decay of the top quarks takes into account 
all spin correlations among the decay products by means of exact 
matrix elements. More information on how to use 
the selection of decay products can 
be found in Appendix B.4. The Higgs boson decay will be soon available.

For the parameterization of the factorization  and 
renormalization scale  the user has three 
choices, {\tt iqopt}=0,1,2, as described in the following Table:
{\renewcommand{\arraystretch}{1.2}
\begin{center}
\begin{tabular}{l||l|l|l}
{\tt iqopt} & 0 & 1  \\  \hline
$Q_0[bb]^2$ & $\hat{s}$ & $\mHsq + \langle \ptsq  \rangle $ & $\langle \ptsq  \rangle $ \\
$Q_0[tt]^2$ & $\hat{s}$ & $(2 m_t + \mH)^2$ & $\langle \ptsq  \rangle $\\
\end{tabular}
\end{center}}
where $\langle \ptsq \rangle $ is the average $\pt^2$ of light and heavy partons. 
Some benchmark results are given in Table~\ref{tab:QQhxs}, obtained
for the following set of cuts and scale choice:
\ba \label{eq:QQh1}
t\tbar : && Q^2=(2 \mt + \mH)^2
\\
b\bbar : && Q^2=\mHsq + (\ptbsq+\ptbbsq)/2 \; .
\label{eq:QQh2}
\ea
 As a default, the Yukawa coupling of the Higgs to the bottom
quarks is evaluated at the b-quark pole mass. Since the rate is
directly proportional to $y_b^2$, the result corresponding to the 
choice $y_b\propto m_b(\mH)$ can be obtained via a trivial rescaling.
The numbers we obtain in the case of 0 extra jets agree with what
 found in the literature~\cite{Carena:2000yx,spira}, after possibly
 correcting for the difference between pole and running $b$-mass.
{\renewcommand{\arraystretch}{1.2}
\begin{table}
\begin{center}
\begin{tabular}{||l|l|l|l||}\hline
$Q \Qbar H$  & $\mH = 120~\gev$  & $\mH = 150~\gev$ & $\mH = 200~\gev$ \\ 
\hline
$Q=t$, LHC (fb) & 401(2) & 212(1) & 89.1(4) \\ 
\hline
$Q=t$, FNAL (fb) & 4.22(2) & 2.00(1) & 0.637(4) \\ 
\hline
$Q=b$, LHC (fb) & 599(3) & 279(3) & 99(2) \\ 
\hline
$Q=b$, FNAL (fb) & 3.73(3) & 1.20(1) & 0.240(2) \\ 
\hline
\end{tabular}            
\ccaption{}{\label{tab:QQhxs} $\sigma(Q \Qbar H)$,
at the Tevatron and 
at the LHC, with parameters given in
eqs.~(\ref{eq:QQh1}-\ref{eq:QQh2}). No cuts applied.}
\end{center}
\end{table}}

\subsection{$\mathbf{ N}$ jets}
\label{sec:Njets}
The subprocesses considered include all configurations with up to two
light quark pairs. They are listed in Table~\ref{tab:Njets}, following
the notations employed in the code. The list covers all possible
processes involving up to 6 light-parton jets with 2 light quark
pairs.  Subprocesses involving 3 light-quark pairs, which only appear
for 4 or more jets, are not included, but are expected to contribute a
negligible rate (they are fully included in the {\tt \small NJETS}
code by Berends et al~\cite{Berends:1989ie}).  In both initial and
final states we only assume as quark types $u$, $d$, $s$ and $c$. For
the generation of events with heavier quarks ($b$ and $t$), we suggest
using the {\tt 2Q} element of the package.

As a default, the code generates kinematical configurations defined by
cuts applied to the following variables:
\begin{itemize}
\item $\pt^{\rm jet}$, $\eta^{\rm jet}$, $\Delta R_{jj} \; .$
\end{itemize}
The respective threshold values can be provided by the user at run
time. Additional cuts can be supplied by the user in an appropriate
routine. 
In the code initialization phase, the user can select 
between 2 choices for the parameterization of the 
factorization and renormalization scale $Q$. A real input
parameter ({\tt qfac}) allows to vary the overall scale of $Q$,
$Q={\tt qfac}\times Q_0$, while the preferred functional form for
$Q_0$ is selected through an integer input parameter ({\tt
iqopt}=0,1).  In more detail:
{\renewcommand{\arraystretch}{1.2}
\begin{center}
\begin{tabular}{l||l|l}
{\tt iqopt} & 0 & 1  \\  \hline
$Q_0^2$ & $\hat{s}$ & $\langle \pt \rangle$ \\
\end{tabular}
\end{center}}
where $\langle \pt \rangle $ is the average $\pt$ of the final state jets. 

\begin{table}
\begin{center}
\begin{tabular}{ll|ll|ll}
{\tt jproc} & subprocess & {\tt jproc} & subprocess & {\tt jproc} &
subprocess \\ 
1 &  $g g \to  gg$ 
&2 &  $q \qbar \to gg$ 
&3 &  $g q \to  qg $ 
\\
4 & $ qg \to qg$
&5 &  $g g  \to   q \qbar$ 
&6 &  $q q   \to qq $ 
\\
7 &  $q q' \to q q'$ 
&8 &  $q\qbar' \to  q \qbar'$ 
&9 &  $q\qbar \to  q \qbar$ 
\end{tabular}
\ccaption{}{\label{tab:Njets} Subprocesses included in the $N$jets
  code. $N-2$ additional final-state gluons are not explicitly 
  shown here but are included in the code, with $N$ up to 6.
  For example, the subprocess {\tt jproc=4},
  in the case of $N=4$ corresponds to the final state  
  $qg \to qggg$.
  The details can be found in the subroutine {\tt selflav} of
  the file {\tt Njetslib/Njets.f}. For each process, the charge-conjugates 
  ones are always understood.}
\end{center}
\end{table}

We cross-checked our calculation against the results for $p\bar{p}\to
Ng$ (with $N\le 6$) given in\cite{Berends:ie}.
Some benchmark results are given in Table~\ref{tab:NJxs}, obtained
for the following set of cuts and scale choice:

\ba
\label{eq:NJscale}
&&Q^2=\langle \pt \rangle\\
&&\pt^{\rm jet}>20~\gev, \quad \vert \eta_j\vert < 2.5, \quad \Delta
        R_{jj} >0.7 \; .
\label{eq:NJcut}
\ea

\begin{table}
\begin{center}
\begin{tabular}{||l|l|l|l|l|l||}\hline
N jets & $N = 2$ & $N = 3$ & $N=4$ & $N=5$ & $N=6$ \\ 
\hline
LHC (nb)  & 375(1)$\cdot 10^3$ & 24.5(2)$\cdot 10^3$ 
& 4,174(8) & 709(1) & 126.3(4)\\ 
\hline
FNAL (nb) &  23,706(60) &  857(5) & 90.9(1) & 8.66(1) & 826(1)$\cdot 10^{-3}$ \\ 
\hline
\end{tabular}            
\ccaption{}{\label{tab:NJxs} $\sigma(N~{\rm jets})$
at the Tevatron and 
at the LHC, with parameters and cuts given in
eqs.~(\ref{eq:NJscale})-(\ref{eq:NJcut}). }
\end{center}
\end{table}

\section{Conclusions}
\label{sec:concl}
We presented in this paper a new MC tool for the generation of
complex, high-multiplicity hard final states in hadronic collisions.
To the best of our knowledge, a large fraction of the processes we
discussed have never been calculated before in the literature to the
level of jet multiplicities considered here, due to the complexity of
the matrix elements involved. In addition to the evaluation of the
matrix elements, and the possibility of performing complete
parton-level simulations, the code we developed offers the possibility
to carry out the shower evolution and hadronization of the partonic
final states. In the current version we implemented the Les Houches
format for the event representation, and developed the relative
interface with \herwig. 
In the future other hard processes (for example including emission of
real, hard photons) will  be added to the list of available
reactions and Higgs decay to two or four fermions will be included.

Our code will allow complete and accurate studies of the SM
backgrounds to a large fraction of the most interesting new physics
phenomena accessible at the Tevatron, at the LHC, and at future
high-energy hadron colliders.

\section*{Acknowledgements}
We thank P.~Richardson and B.~Webber for invaluable help in the
development of the Les Houches compliant \herwig\ interface for
\herwig, and A.~Messina for his contribution to the \pythia\ interface.
FP thanks G. Montagna and O. Nicrosini for their collaboration during
the early stage of the development of the $Q\Qbar Q'\Qbar'$ processes,
and the Pavia Gruppo IV of INFN for access to the local computing resources.

\begin{appendix}
\section{The contents of the code package}
The code is written in Fortran77, with the part relative to the matrix
  element evaluation available as well in Fortran 90 (see
  Appendix~\ref{sec:f90}).
The code package, contained in the compressed file {\tt alpgen.tar.gz}, 
can be obtained from the URL
{\tt  http://home.cern.ch/mlm/alpgen}.

Unpacking the zipped tarred file {\tt alpgen.tar.gz} with the command:

\begin{verbatim}
> tar -zxvf alpgen.tar.gz
\end{verbatim}
will create the following directory structure:

\begin{verbatim}
2Qlib/     Njetswork/  herlib/     wcjetwork/  zjetlib/
2Qwork/    QQhlib/     pylib/      wjetlib/    zjetwork/
4Qlib/     QQhwork/    vbjetslib/  wjetwork/   zqqlib/
4Qwork/    VF90/       vbjetswork/ wqqlib/     zqqwork/
Njetslib/  alplib/     wcjetlib/   wqqwork/
\end{verbatim}

The directories labeled {\tt 'lib'} contain the source codes for the
respective processes. The user is not supposed to touch them. 
The code elements which the user will need to access and possibly
modify to run his own analyses are contained in the directories
labeled {\tt 'work'}. More in detail:

\begin{itemize}
\item The directory {\tt alplib/} contains the parts of code which are
  generic to the evaluation of matrix elements using the \ALPHA\ 
  algorithm.  The user should treat this directory as a black box.
  When new processes will be added in the future, this part of the
  code should not change.  More in detail:
  \begin{itemize}
  \item {\tt alplib/alpgen.f}: contains the general structure of the
    code, preparing the input for the matrix element calculation, the
    bookkeeping of the cross-section determination, the event
    generation, etc.
  \item {\tt alplib/alpgen.inc}: include file, with the necessary common
    blocks.
  \item {\tt alplib/Aint.f, Asu3.f, Acp.f}: the set of programmes
    necessary for the calculations of the matrix element, done by the
    \ALPHA\  algorithm.
  \item {\tt alplib/alppdf.f}: contains a collection of structure
    function parameterizations; some of them require at run time input
    tables, which are provided as part of the package, and stored in
    the subdirectory {\tt alplib/pdfdat/}. The command file {\tt
      alplib/pdfdat/hvqpdf} contains the necessary logical links to
    all PDF data tables. As a default, we already provide a logical
    link to this file in all {\tt /*work} directories. It is
    sufficient to issue the {\tt pdflnk} command within the desired
    working subdirectory to create the necessary logical links, and
    allow the use of all available PDFs.
  \item {\tt alplib/alputi.f}: This program unit contains a
    histogramming package which allows to generate
    {\tt topdrawer}~\cite{topdrawer}  files with
    the required distributions. Examples of the use of this package
    are provided in the default user files {\tt *work/*usr.f}.  
    Users who prefer other histogramming packages, such as HBOOK, do
    not need to link to this file.
  \end{itemize}
  
\item The directories {\tt *lib/} ({\tt *=wqq, zqq, wcjet, wjet, zjet,
    vbjets, 2Q, 4Q, QQh, Njets}) contain the parts of the code
    specific to the generation of $WQ\Qbar+$~jets, $ZQ\Qbar+$~jets,
    $W+$~jets, $W+c+$~jets, $Z+$~jets, $nW+mZ+lH+$~jets,
    $Q\Qbar+$~jets, $Q\Qbar Q'\Qbar'+$~jets $Q\Qbar H+$~jets and
    $N$~jets events.  The respective {\tt include} files with the
    necessary process-dependent common blocks are included in these
    directories.  The user should treat these directories as black
    boxes.
  
\item The directories {\tt *work/} ({\tt  *=wqq, zqq, wcjet, wjet, zjet,
    vbjets, 2Q, 4Q, QQh, Njets}) contain the parts of the code which the user
  is supposed to interact with, in order to implement his own analysis
  cuts, etc.  They contain the files {\tt *usr.f}, where sample
  analysis routines are provided. These files host the routines in
  which the user can select generation mode, generation parameters
  (e.g. beam energy, PDF sets, heavy quark mass, etc.) as well as
  generation cuts (minimum \pt\ thresholds, etc.). Here the user
  initialises the histograms, writes the analysis routine, and prints
  out the required program output.  This is the only part of the code
  in which the user is supposed to operate, editing the analyses
  files, and producing and running the executable
  (see next Section). The versions provided as a
  default contain already complete running examples, with the
  respective command files ({\tt input}) containing sets of default
  settings (see next Section for more details on running the code).
    
\item The directory {\tt herlib/} contains the parts of code relevant
  for the shower evolution using \herwig. In addition to the \herwig\ 
  source and include files for version 6.4, this directory includes
  the file {\tt atoher.f}, which is the interface between the
  parton-level matrix elements and \herwig, and the file {\tt
    hwuser.f}, which includes the main driver for the running of \herwig,
  and the part of the code where the user can input the analyses
  routines. More in detail:
  \begin{itemize}
  \item {\tt hwuser.f}: user initialization of the analysis. Includes
    the standard \herwig\ initialization, histogram intialization,
    analysis routines, etc.  The calls to new routines {\tt hwigup}
    and {\tt hwupro} create the interface with the generated hard
    events. 
  \item {\tt atoher.f}: this file contains all routines necessary to
    read in the unweighted events produced by the hard matrix element
    generator.  The routine {\tt hwigup} downloads the initialization
    parameters of the hard process (process type, number of partons,
    beam energy and beam type, etc.), and allows the main \herwig\
    initialization. The routine {\tt hwupro} is called for each event:
    it reads the event kinematics, flavour and colour information from
    the file of unweighted events, and translates the event data to
    allow the \herwig\ processing of the shower.  This file should be
    treated by the user as a black box.
  \item {\tt herwig64.f}: the {\tt herwig64} source code. 
  \item {\tt HERWIG64.INC, herwig6400.inc}: \herwig\  common blocks
  \item {\tt pdfdummy.f}: dummy PDF routine, required by \herwig\ 
    unless the CERN library PDF sets are used. As a default, the
    current version runs with the default \herwig\ PDF set, regardless
    of the PDF set which was used to generate the hard process. We
    verified that this does not affect the features of the showered
    final state.  Nevertheless we plan in the future to enforce
    the consistency between PDF set used in the hard generation and in
    the shower evolution.
  \end{itemize} 
\item A similar directory and file structure is provided for \pythia.
\end{itemize} 

\section{Running the code}
To compile the code for the $WQ\Qbar$ process\footnote{Analogous
  procedures allow compilation and run of other processes.}, change
directory to:

\begin{center}  {\tt > cd wqqwork}  \end{center}
A {\tt Makefile} is provided for compilation. Issue the comand 

\begin{center}  {\tt > make wqqgen}  \end{center}
and the executable {\tt wqqgen} will be prepared. It can be run
interactively, inputting from the keyboard the run parameters
requested by the code, or using
the default command file {\tt input}, issuing the command 

\begin{center}  {\tt > wqqgen < input}  \end{center}
Editing the file {\tt input} allows to change the initialization
defaults (e.g. the number of jets, the heavy quark masses, the PDF
sets, etc.). Notice that
altering some of the inputs in the {\tt input} file may influence the
sequence of parameters requested from the code; for example,
depending on whether light jets are requested or not in a given
process, the generation cuts for these jets will or will not be
required in input. The {\tt input} file should therefore be edited in
a consistent manner.

The first input parameter requested is the running mode {\tt
  imode}. The three available running modes are discussed in detail in
the following subsections. 

The codes {\tt vbjets}, {\tt 2Q} and {\tt QQh} include the option 
{\tt idecay}, to allow for the decay of the generated on shell 
vector bosons and top quarks, taking into account spin correlations 
between fermion decay products by means of decay matrix elements. 
More details on how to run the codes are discussed in the dedicated subsection.

\subsection{{\tt imode=0}}
The simplest option is {\tt imode=0}, where events are generated
according to the selected cuts, a total cross section is evaluated,
and the user can use the routine {\tt evtana} to analyse the event and
fill histograms with the desired distributions. To facilitate the job
of the user, we provide a (redundant) array of kinematical variables
relative to the event. The array is initialised in the routine {\tt
  usrfll} contained in the file {\tt wqqlib/wqq.f}, and is 
stored in the common block 
{\tt usrevt} contained in the include file {\tt wqq.inc}. Examples of
variables provided include:
\begin{itemize}
\item {\tt pin(4,2):} momenta of the incoming partons
\item {\tt pout(4,8):} momenta of the outgoing particles (maximum 8
  outgoing partons)
\item {\tt pjet(4,8):} momenta of the final-state partons
  (i.e. quarks and gluons)
\item {\tt ptj(8), etaj(8):} transverse momentum and pseudorapidity
  of the final-state partons
\item {\tt pbott(4) (pbbar(4)), ptb (ptbb):} momentum and transverse
  momentum of the heavy (anti)quark 
\item {\tt plep(4) (ptlep), pnu(4) (ptmiss):} (transverse) momentum 
 of the charged lepton and neutrino
\item {\tt drjj(8,8):}  $\Delta R$ separation in $\eta-\phi$ space
  among the final-state partons
\item {\tt drbj(8) (drbbj(8)) :}  $\Delta R$ separation in $\eta-\phi$ space
  between the heavy (anti)quark, and the final-state partons
\item etc\dots
\end{itemize}
Similar sets of variables are provided for the other processes.  As an
output the user will find the following files: ({\tt `file'} is the
label assigned by the user at run-start time):
\begin{itemize}
\item {\tt file.stat}: the header of this file contains information on
  the run: value of the input parameters (EW and strong couplings,
  beam types and energies, PDF set), hard process selected and
  generation cuts. Furthermore, this file reports the results of each
  individual integration cycle, with total cross sections, as well as
  individual contributions from the allowed subprocesses. It gives the
  maximum weights of the various iterations, and the corresponding
  unweighting efficiencies, and the value of the cross-section
  accumulated over the various iterations, weighted by the respective
  statistical errors.
\item {\tt file.top}: includes the topdrawer plots of the
  distributions, if requested. The default normalization of the
  histograms is in pb/bin.
\item {\tt cnfg.dat}: file required by ALPHA, generated at run time;
         it is not needed for the
         analysis of the output, and will be recreated anew any time
         the code runs, so the user should not bother about it, and
         it can be safely deleted at any time.
\item {\tt file.mon}: produced/updated after each 100K events;
       it  contains information on the status of the
         run, dumped  every 100K events. It is useful to monitor the
         progress  of the run. In addition to this monitoring tool,
         the user can choose to perform other tests during the run, in
         order to save partial information, or monitor the evolution
         of the plotted distributions. Every 100K events the program
         calls the routine {\tt monitor}, contained in the user file {\tt
           wqqwork/wqqusr.f}, where the user can select which
         operations to perform. As a default, the provided routine
         prints out each 1M events the topdrawer file with the
         distributions being histogrammed. In case of crash, the
         results relative to the statistics accumulated up to that
         point  are therefore retrievable.
\item {\tt file.grid*}: The phase-space is discretised and
         parameterised by a multi-dimensional grid. During the
         phase-space integration, a record is kept of the rate
         accumulated within each bin of each integration variable. At
         the end of an integration cycle (``iteration''), the
         total bin-by-bin rates are used to improve the grid
         sampling efficiency. This is achieved by assigning sampling
         probabilities proportional to the bin integrals (we ensure
         however that 20\% of the sampling is uniformly distributed
         among all bins, to avoid artificial biases introduced by runs
         with limited statistics). A subsequent iteration can then
         benefit from a better sampling. The state of the grid at the
         end of each iteration is saved in the file {\tt file.grid1}.
         Since the first few
         iterations give rise to distributions which are likely to be
         biased by large statistical fluctuations, we separate a phase
         of grid warm-up from a phase in which events will be
         generated and distributions calculated. The user should then
         specify in the {\tt input} file the number of warm-up
         iterations, the number of events to be calculated for each
         iteration, and then the number of events that will be used
         for the final event generation and for the analysis. At the
         end of the event generation, the grid will be saved to the
         file {\tt file.grid2}. We keep this separate from {\tt
           file.grid1} to allow the user to choose whether to start a
         new generation cycle
         using the grid status at the end of the previous warm-up
         phase, or at the end of the previous generation phase.
         These choices are made by the user in the {\tt input} file,
         selecting the variable {\tt igrid} to be 0 (to reset the
         grids and start a new grid optimization), 1 (to start the new
         run using the grid obtained during the previous warm-up
         phase) or 2 (to start the new run using the grid optimised at
         the end of the latest event generation). Both grid files are
         saved in {\tt file.grid*-old} (*=1,2) at the beginning
         of each run, to allow recovery of the grid information in
         case of run crashes or mistakes.
\end{itemize}
\subsection{{\tt imode=1}}
Running the code with {\tt imode=1} offers the same functionality than
{\tt imode=0}, but will in addition write the weighted events to a
file. To limit the size of the file, only events which passed a
pre-unweighting are saved. The pre-unweighting is based on a maximum
weight $w_{tmp}$ which is equal to 1\% of the actual maximum weight at
the moment of the generation of the event: $w_{tmp}=w_{max}/100$.  An
event with weight $w$ passing the pre-unweighting is then assigned a
weight $w'=w_{tmp}$ if $w<w_{tmp}$, or $w'=w$ if $w>w_{tmp}$. The
weight $w'$ is then saved to a file, together with the random number
seed which initiated the generation of this event, and with the value
of $x_1$ (useful to check the sanity of the file when it will be read
again for the unweighting).  Some statistical information on the run,
including the total number of generated events, the integral, and the
overall maximum weight, are saved as well in a separate file. The file
with weighted events is to be used for a later unweighting. One can
easily verify that the pre-unweighting procedure does not introduce
any bias in the final unweighting. The random number seed will then be
sufficient to regenerate the full kinematical, flavour and colour
information on the event. The size of each event is 57~bytes. Make sure
you have enough disk space to write out the number of events you
require.  In addition to the files listed above for {\tt imode=0}, as
an output the user will find the following files:
\begin{itemize}
\item {\tt file.par}: includes run parameters (e.g. beam energies and types,
  generation cuts, etc), phase-space grids,  cross-section and maximum-weight
  information;
\item {\tt file.wgt}: for each event we store the two seeds of the
  random number generation, the event weight, and the value of $x_1$
  for the event (as a sanity benchmark after the kinematics has been
  reconstructed from the random seeds)
\end{itemize}
As a default, the bookkeeping of the weight distribution is kept in
the routine {\tt *usr.f}, and the relative data are printed in the
topdrawer file {\tt file.top}. The study of the weight distribution
can guide the user to a more efficient choice of maximum weight before
starting the event unweighting.
\subsection{{\tt imode=2}}
After a run with {\tt imode=1}, a run with {\tt imode=2} will perform
the unweighting of the already generated events, and will prepare the
input file for the processing of the events with \herwig~or~\pythia.  
The code
reads first the phase-space grids used for the weighted-event
generation and
 the maximum weight from {\tt file.par}. The user has the possibility
 to edit {\tt file.par} and replace the maximum weight with a
 different value, if he convinces himself that a more efficient
 unweighting can be obtained, without biasing the sample, by selecting a
 smaller maximum weight\footnote{The user must however avoid
  using a maximum weight smaller than 1\% of
  the true maximum weight, because of the threshold used in 
the pre-unweighting phase.}. We are working on techniques to perform
this optimised unweigthing in an automatic way. The code will then scan the
file {\tt file.wgt}, containing the weighted events. A comparison of
the event weights with the maximum weight is made, and the unweighting
is performed. The kinematics of each unweighted event is reconstructed
from the relative random number seed. The colour flow for the event is
then calculated, and the full event information is written to a new
file. This file will be the starting point for the generation of the
full shower, to be performed using \herwig\ or \pythia.  As an output the user will
find the following files:
\begin{itemize}
\item {\tt file\_unw.stat}: includes cross sections, max weight, etc;
\item {\tt file\_unw.top}: While no new events are generated, the analysis
  routines used when running in {\tt imode=1} are applied to the
  unweighted events, and the relative distributions are evaluated. In
  this way the user can compare distributions before and after unweighting;
\item {\tt file.unw}: list of unweighted events, including event
  kinematics, flavour and colour structure, and event weight.
\end{itemize}

\subsection{Decay of top quarks and vector bosons with spin correlations}
The on-shell top quarks generated by {\tt 2Q} and {\tt QQh}  undergo 
a fully exclusive decay in three fermions weighted with exact matrix 
element.
When running in {\tt imode=0,1} the information on top decay product 
momenta is stored in the matrix {\tt idec(4,3,2)}, available in 
the routines {\tt usrfll} and {\tt usrcut}. The meaning of the entries 
is as follows: {\tt idec(1:4,i,j)} is the four momentum ($p_x, p_y,
p_z, E$) 
of the i-th decay product of the j-th particle ({\tt jtl} and {\tt jtbl} 
are the labels for top and antitop quark respectively); 
{\tt i=1} is the label for the $b$ ($\bar{b}$) quark; {\tt i=2,3} 
are the labels for the fermion and antifermion coming from the $W$ decay 
respectively. When running in {\tt imode=2} the user is required
(interactively) to 
select one top decay mode among seven options: 
\begin{eqnarray}
{\tt 1} &=& e \nu_e b \bar b + 2\, {\rm jets},  \nonumber \\ 
{\tt 2} &=& \mu \nu_\mu b \bar b + 2\, {\rm jets}, \nonumber \\
{\tt 3} &=& \tau \nu_\tau b \bar b + 2\, {\rm jets},  \nonumber \\
{\tt 4} &=& l \nu_l b \bar b + 2\, {\rm jets},\, \, 
(l = e, \mu, \tau) \nonumber  \\
{\tt 5} &=& l \nu_l l' \nu_{l'} b \bar b, (l = e, \mu, \tau) \nonumber  \\
{\tt 6} &=& b \bar b + 4\,{\rm jets}, \nonumber \\
{\tt 7} &=& {\rm fully} \,\, {\rm inclusive}. \nonumber 
\end{eqnarray}
Non-zero  masses of the $W$-decay products are introduced by rescaling 
their momenta, and keeping invariant the $W$ 4-momentum.

The on-shell $W$ and $Z$ gauge bosons generated in {\tt vbjets} undergo 
a fully exclusive decay. 
While the decay products of the $W$ can be changed when running in 
{\tt imode=2} (see below), given the universal electroweak couplings of 
the $W$ to 
fermions, the $Z$ decay options must be specified at the very beginning. 
When running in {\tt imode=0,1} the information on the $W$, $Z$ decay product 
momenta is stored in the matrix {\tt idec(4,4,maxpar-2)}, available in 
the routines {\tt usrfll} and {\tt usrcut}. The meaning of the entries 
is as follows: {\tt idec(1:4,i,j)} is the four momentum ($p_x, p_y,
p_z, E$) 
of the i-th decay product of the j-th particle; {\tt i=1,2} are the labels 
for fermion and antifermion respectively. The ordering in {\tt j}
correspond to the ordering in which $Z$ and $W$'s are generated.
 The flavour of the $Z$ decay products is stored 
in the variable {\tt zfl(maxpar-2)} according to PDG conventions. 
For every $Z$ boson in the final state the user have to select its decay 
mode in the input file by entering an integer string with the decay modes 
of the individual $Z$'s according to the following table
\begin{eqnarray}
{\tt 1} &=& \nu {\bar \nu}, ({\rm summed}\, \, {\rm over }\, \, {\rm all}\, \, 
{\rm flavours}), \nonumber \\
{\tt 2} &=& l^+ l^-, ({\rm summed}\, \, {\rm over }\, \, {\rm all}\, \, 
{\rm flavours}), \nonumber \\
{\tt 3} &=& q {\bar q}, ({\rm summed}\, \, {\rm over }\, \, {\rm all}\, \,
{\rm flavours}), \nonumber \\
{\tt 4} &=& b {\bar b}, \nonumber \\
{\tt 5} &=& {\rm fully} \,\, {\rm inclusive}. \nonumber 
\end{eqnarray}
Concerning the decay modes of the $W$'s, they have to be specified 
when running in {\tt imode=2} as follows (the code asks for this
automatically and interactively):
\begin{eqnarray}
{\tt 1} &=& e {\bar \nu_e}, \nonumber \\
{\tt 2} &=& \mu {\bar \nu_\mu}, \nonumber \\
{\tt 3} &=& \tau {\bar \nu_\tau}, \nonumber \\
{\tt 4} &=& l {\bar \nu}_l (l = e, \mu, \tau),  \nonumber  \\
{\tt 4} &=& q {\bar q}',  \nonumber  \\
{\tt 5} &=& {\rm fully} \,\, {\rm inclusive}. \nonumber 
\end{eqnarray}
The same final state options for $W$ bosons are available also 
for {\tt wjets} and {\tt wqq} with {\tt imode=2}.
Finite fermionic masses of the decay products are introduced by rescaling 
the momenta and preserving the vector boson 4-momentum.

\subsection{Running \herwig\ or \pythia\ on the unweighted events}
A \herwig\ executable can be obtained starting
from the default driver {\tt herlib/hwuser.f}. To compile and link,
issue the command:
\begin{center}  {\tt > make hwuser}  \end{center}
from the {\tt herlib} directory. The resulting executable, {\tt
  hwuser}, should then be run in the directory containing the
unweighted event file. The default code simply generates the shower
evolution for the given events, and prints to the screen the particle
content of the first few events (number of printed events
set by the variable {\tt maxpr}). Analysis code should be provided by
the user, by filling the standard \herwig\ routines {\tt hwabeg,
  hwanal, hwaend}. A log file {\tt file-her.log} documenting the
inputs and outputs of the run is produced.  An analogous interface for
running \pythia\ is available.

\section{Portability and  Fortran 90}
\label{sec:f90}
The code was tested on several platforms, including Linux based PC's,
Digital Alpha Unix, HP series 9000/700, Sun work stations and MAC-OSX
with a {\tt g77} compiler.
The user should however check that the compilation
options provided by default in the first few lines of the {\tt
  Makefile} files, including the choice of Fortran compiler, are
consistent with what he has available. We shall be happy to receive
comments related to the portability of the code, and will update the
code to improve its usability.

Toghether with the old Fortran77 ({\tt F77}) version of the \ALPHA\ code
\cite{Caravaglios:1995cd,Caravaglios:1999yr}, we provide a new 
  Fortran90 ({\tt F90}) version.  The evaluation of the \ALPHA\ matrix elements
with the {\tt F90} version is a factor of five to twenty times faster
than the {\tt F77} version, depending on the selected process (the
more complex the process, the better the improvement). When the
overhead of the rest of the code (phase-space, parton densities, etc)
is added in, the overall performance of the code improves by a factor
of two to five (we stress that only the \ALPHA\ part of the code is
available in {\tt F90}; users unfamiliar with {\tt F90} should not be
discouraged from using this version, since this component is a black
box, and its use is compatible with the {\tt F77} part of the code
which the user has access to. Furthermore, the {\tt F90} executables
will run using the same {\tt input} files as the {\tt F77} versions of
the code, and produce the same results, to machine precision).

To link the F90 version of \ALPHA\ it is sufficient to input the choice
of {\tt F90} compiler in the {\tt Makefile}, and issue the comand
{\tt make wqqgen90}, which will produce the executable {\tt wqqgen90}.

For user who do not have access to a {\tt F90} compiler, we provide
one suitable for running on PC's with Linux operating
systems. To set it up, proceed as follows: 
\begin{itemize}
\item go to the {\tt ALPGEN} home directory;
\item issue the command {\tt make ft90V}. This will unpack the file
  {\tt ft90V.tar.gz} and install the {\em Vast/Verydian {\tt F90}}
    compiler into the directory {\tt F90V}. This software was
  distributed freeware for personal use only by Pacific-Sierra
  Research. Before use, you are therefore supposed to agree with the license
  term contained in the directory {\tt F90V/}. In the same directory
  the user can find some documentation on the compiler, including the
  list of supported platforms.
\item Move to the desired work directory (e.g. {\tt wqqwork});
\item issue the command {\tt make wqqgen90V} which will produce the executable
{\tt wqqgen90V}. 
\end{itemize}


\end{appendix}

\end{document}